\pdfoutput=1

\documentclass[reprint,aps,prl,showpacs,twocolumn]{revtex4-1}
\usepackage{mathptmx}
\usepackage{epsfig,amsopn}
\usepackage{graphicx}
\usepackage{amsmath,amssymb}
\usepackage{array} 
\usepackage{tabularx}
\usepackage{booktabs}
\usepackage[dvipsnames]{xcolor}
\usepackage{braket}
\usepackage{float}
\usepackage{enumerate}
\usepackage{mathtools}
\usepackage[symbol]{footmisc}
\allowdisplaybreaks[1]
\usepackage{wrapfig}
\usepackage{makecell}
\usepackage{enumitem}
\usepackage{soul}
\usepackage{multirow}
\usepackage{etoolbox}

\usepackage[most]{tcolorbox}
\usepackage{booktabs}
\usepackage{tikz}
\usetikzlibrary{arrows.meta}
\usetikzlibrary{shapes.arrows}


\def\bea{\begin{eqnarray}}
\def\eea{\end{eqnarray}}

\usepackage{hyperref, url}
\usepackage{stmaryrd}

\usepackage[capitalise]{cleveref} 

\begin{document}

\title{First Passage in the Presence of Jump Rate Fluctuations:\\[6pt]  A Transition from Infinite to Finite Mean}

\author{{\normalsize{}Bara Levit$^1$}{\normalsize{}}}

\author{{\normalsize{}Shlomi Reuveni$^1$}{\normalsize{}}}
\email{Author for correspondence: shlomire@tauex.tau.ac.il}

\affiliation{\noindent \textit{
$^{1}$School of Chemistry, The Center for Computational Molecular and Materials Science, The Center for Physics and Chemistry of Living Systems, Tel Aviv University, Tel Aviv, Israel}}

\date{\today}

\begin{abstract}
\noindent
A simple symmetric random walk on a one-dimensional lattice is guaranteed to reach any target site, yet its mean first-passage time diverges. Here, we show that stochastic fluctuations in the jump rate can fundamentally alter this classic result. We find a phase transition in the mean first-passage time: sufficiently broad rate fluctuations render it finite, whereas weaker fluctuations leave it infinite. Our results demonstrate that temporal fluctuations in the jump rate---and hence in the diffusivity---can overcome spatial wandering and regularize otherwise divergent first-passage times.
\end{abstract}

\maketitle

Flip a coin to pick your direction, take a step, and repeat. This deceptively simple rule defines the random walk. Its reach is astonishing: it traces the jittering path of pollen in water \cite{Brown}, guides animals searching for food \cite{Search_Processes,Animal_Search_Strat,sims2019optimal,benichou2011intermittent,vilk2022ergodicity}, and drives the algorithms behind AI image generators \cite{Diffusion_Model_AI, AI_noise_SDE_diffusion,Deep_Unsup_Learning_Diffusion,biroli2024dynamical}. Its simplest incarnation, the one-dimensional simple symmetric random walk, is perhaps the most widely studied model of stochastic motion. One of its most celebrated properties is recurrence: every site is eventually visited with probability one. Yet this guarantee of success hides a striking pathology. While the probability to reach any site is unity \cite{polya1921aufgabe}, the average time to do so — the mean first-passage time (MFPT) — diverges \cite{redner2001guide,klafter2011first,bray2013persistence,Kostinski_infinitemean}.

In the simple random walk, time only advances when a step is taken, so it is naturally discrete. A standard generalization replaces these unit time steps by random waiting times, yielding the continuous-time random walk (CTRW) \cite{CTRW_MontrollWeiss,  Shlesinger1973, CTRW_FokkerPlanck, CTRW_Ergodic, Scher2017, CTRW_SokolovKlafter, Shlesinger2017}. Yet, this modification does not regularize the MFPT, which remains divergent. The reason is that the divergence is fundamentally spatial rather than temporal. It is driven by rare but arbitrarily long excursions away from the target, whose contribution overwhelms any choice of waiting-time statistics.

In the simplest version of the CTRW, waiting times between steps are governed by a constant jump rate. The jump rate sets the pace at which the walker explores its surroundings, mirroring the role of the diffusion coefficient. Recent studies randomize the diffusion coefficient itself \cite{DiffusingDiffusivitySlater, DD_Superstatistics} to explain how particles have a non-Gaussian displacement distribution while retaining a mean-square displacement that grows linearly with time \cite{DD_Experiments1, DD_Experiments2}. Fluctuations in diffusivity create periods of both slow and rapid motion. Low-diffusivity periods can temporarily `trap' the particle near its current position, whereas high-diffusivity episodes enable it to explore space rapidly and increase the probability of reaching a distant target at shorter times \cite{DD_FPT,DD_FPT2,Stochastic_Diffusion_FPT,Grebenkov_NatureCom,sposini2024being-a,sposini2024being-b}.

The altered first-passage behavior produced by fluctuating diffusivity invites a closer look at its microscopic origin. Temporal variations in diffusivity naturally correspond to fluctuations in the rate at which steps occur. Previously, this motivated extending the standard CTRW by treating the jump rate itself as a stochastic process. This gave rise to the Doubly Stochastic CTRW (DSCTRW) \cite{arutkin2023doubly}. Here, we consider the impact of such jump-rate fluctuations on the first-passage behavior of the simple symmetric random walk. We show that they induce a remarkable phase transition: beyond a critical strength of the fluctuations, the MFPT changes from infinite to finite. This uncovers a new route by which temporal fluctuations qualitatively modify transport and search processes.

\begin{figure}[t]
\centering
\includegraphics[width=0.485\textwidth]{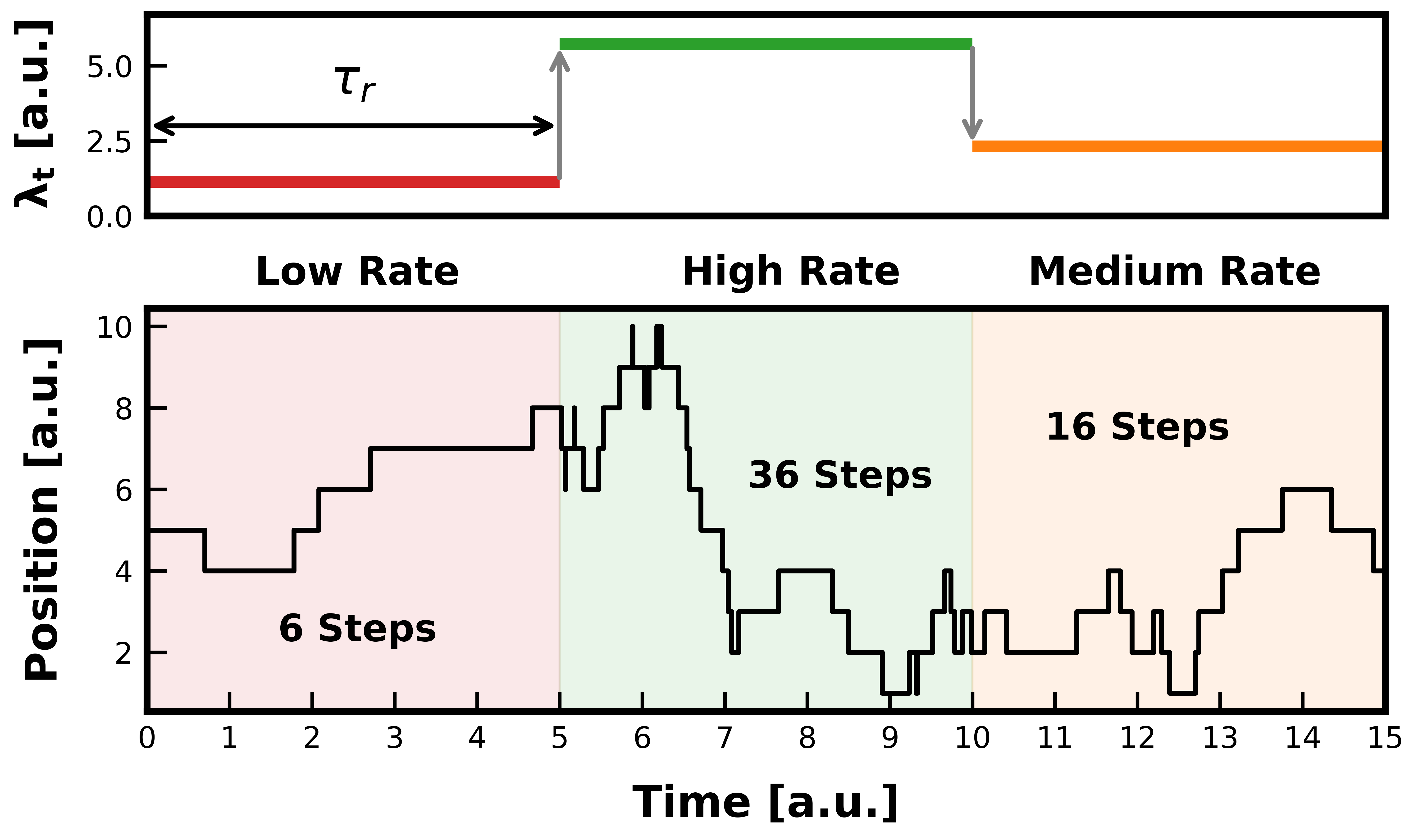}
  \caption{\textbf{Example of a doubly stochastic continuous-time random walk (DSCTRW)}. The top and bottom panels show the rate and position as a function of time, respectively. In this particular example, the jump rate is resampled, from an exponential distribution with unit mean, at fixed time intervals of length $\tau_r=5$ [a.u.].}
  \label{fig:Demo_Walk}
\end{figure}

\begin{figure*}[t]
\centering
\includegraphics[width=0.95\textwidth]{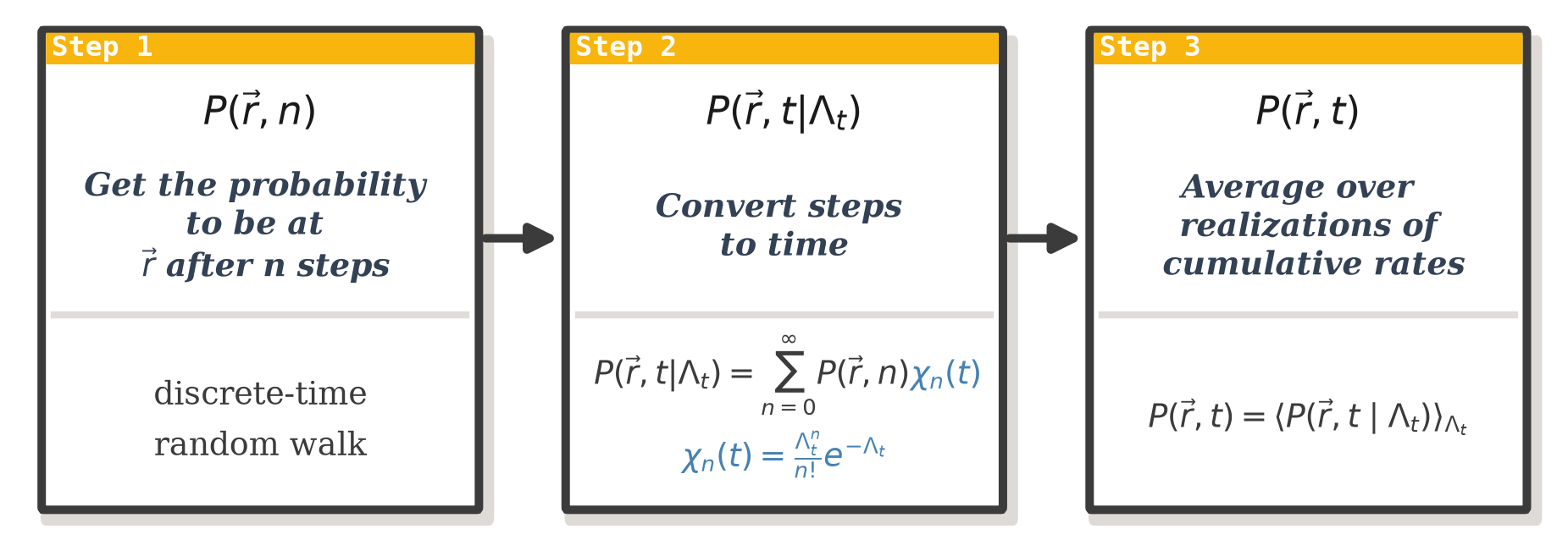}
 \caption{\textbf{DSCTRW Recipe}.  Follow these steps to obtain the probability to be at position $\vec{r}$ at time $t$ in the DSCTRW. Step 1: Start with the probability to be at $\vec{r}$ after $n$ steps for the discrete-time random walk. Step 2: Convert steps to time: Multiply the probability in step 1 by $\chi_n(t)$, the probability to take $n$ steps by time $t$ given a specific realization of the cumulative jump
rate $\Lambda_t$. This is a Poisson distribution with mean $\Lambda_t$. Then, sum over all steps. Step 3: Average over all possible cumulative rates $\Lambda_t$.}
  \label{fig:Recipe}
\end{figure*}

\textit{The DSCTRW.---}We begin by briefly recalling the DSCTRW, the minimal framework that captures temporal fluctuations in the jump rate of a random walk. Consider the Montroll-Weiss continuous-time random walk (CTRW) \cite{CTRW_MontrollWeiss, Scher2017, CTRW_SokolovKlafter, Shlesinger2017}. In the simplest setting, steps occur at a constant rate \(\lambda\), such that the time between successive steps is exponentially distributed. This is the first layer of stochasticity. To obtain the DSCTRW, we introduce a second layer of randomness: a time-dependent jump rate, \(\lambda_t\), that is itself a stochastic process. In this picture, the probability to make a jump during a time interval $[t,t+dt]$ fluctuates in time and is given by $\lambda_t dt$. For a given realization of $\lambda_t$, steps in the DSCTRW follow  an inhomogeneous Poisson process. The number of steps in the time interval $[0,t]$ is thus a Poisson random variable with a mean that equals the cumulative jump rate $\Lambda_t\equiv\int^t_0 \lambda_{t^\prime} dt^\prime$.

To illustrate the DSCTRW, we consider a minimalistic example: a simple symmetric walk on a one-dimensional discrete lattice. Moreover, in this example, the jump rate is resampled every fixed time interval of duration $\tau_r$, from some distribution. Thus, $\lambda_t$ becomes a piecewise constant function. As a consequence, within every time interval the walk  behaves as a standard CTRW with a constant jump rate. We interpret $\tau_r$ as the relaxation time, since it sets the characteristic timescale beyond which previous values of $\lambda_t$ are forgotten.

\cref{fig:Demo_Walk} illustrates the resulting walk. The top panel shows how the rate changes with respect to time. The rate starts low (red, first time interval), changes to high (green, second time interval), and finally drops to medium (orange, third time interval). The bottom panel shows the  position of the walk as a function of time. A step changes the position by $\pm 1$ (vertical lines); otherwise the position is unchanged (horizontal lines). The first, second, and third intervals contain six, thirty-six, and sixteen steps, respectively. Observe that when the rate is higher more steps are taken and vice versa. 

How does one solve for this walk? We will present here a ‘recipe’, applicable for any walk and stochastic jump rate $\lambda_t$. The recipe consists of three steps and is showcased in \cref{fig:Recipe}. Start with $P(\vec{r},n)$, the probability to be at position $\vec{r}$ after $n$ steps, for a general random walk in discrete time. Next, similar to the  CTRW, convert steps to time. Namely, multiply $P(\vec{r},n)$ by $\chi_n(t)\equiv \frac{\Lambda_t^n}{n!}  e^{-\Lambda_t}$, the probability to take exactly $n$ steps by time $t$ for a given realization of the cumulative jump rate $\Lambda_t$. In the case where the jump rate is resampled every $\tau_r$,  $\Lambda_t = \overset{n_{\tau_r}-1}{\underset{i=0}{\sum}} \tau_r \lambda_{i\cdot \tau_r} + \delta_t \lambda_{n_{\tau_r}\cdot \tau_r}$, where $n_{\tau_r}=\left\lfloor \frac{t}{\tau_r}\right \rfloor$ is the number of complete $\tau_r$ intervals in $[0,t]$, $\lambda_{i\cdot \tau_r}$ is the value of $\lambda_t$ in the time interval $t\in[i\cdot \tau_r,(i+1)\cdot \tau_r)$, and $\delta_t = t-n_{\tau_r}\tau_r$ is the leftover, or residue time (Supplementary \cref{sec:Piecewise_Stochastic_Rate}). In the example of \cref{fig:Demo_Walk}, $\tau_r=5$ [a.u.] and for $t=15$ [a.u.] we have $n_{\tau_r}=3$ and $\delta_t=0$. After converting steps to time we obtain $P(\vec{r}, t| \Lambda_t)$, the probability to be at $\vec{r}$ at time $t$, given $\Lambda_t$.

To finish the recipe, proceed to the third step, which incorporates the second layer of stochasticity, i.e., average over all stochastic realizations of $\lambda_t$, and hence over $\Lambda_t$. This gives $P(\vec{r},t)$, the probability to be at $\vec{r}$ at time $t$ in the DSCTRW. In Fourier space the above recipe gives particularly elegant expressions. Indeed, the Fourier transform of $P(\vec{r},t|\Lambda_t)$ is given by $\hat{P}(k,t|\Lambda_t)=\overset{\infty}{\underset{n=0
}{\sum}} \chi_n(t)\hat{\phi}(k)^n = e^{-(1-\hat{\phi}(k))\Lambda_t}$, where $\hat{\phi}(k)$ is the characteristic function of the single step distribution governing the discrete-time random walk. Averaging over all realizations of cumulative jump rates, we obtain that the Fourier transform of $P(\vec{r},t)$ is $\hat{P}(k,t) = \tilde{\Lambda}_t(1-\hat{\phi}(k))$, where $\tilde{\Lambda}_t(1-\hat{\phi}(k))$ is the Laplace transform of $\Lambda_t$ evaluated at $1-\hat{\phi}(k)$. This form highlights that the single-step characteristic function, $\hat{\phi}(k)$, and the Laplace transform of the cumulative rate, $\tilde{\Lambda}_t$, are enough to determine the position distribution in the DSCTRW.

\label{Section:Half-Line} \textit{First Passage on the Half-Line.---}The goal of this letter is to understand how stochastic fluctuations in the jump rate affect the resulting first-passage statistics. We consider a simple symmetric walk on the one-dimensional lattice, in which the jump rate is resampled every relaxation time $\tau_r$. The walker starts at position $x_0>0$, while the origin ($x=0$) acts as an absorbing boundary. The process terminates upon the walker's first arrival to the origin. 

In ref.~\cite{arutkin2023doubly}, the jump rate was drawn from an exponential distribution. Here we generalize this choice and draw the rate from a Mittag-Leffler distribution: $g(\lambda)=\overset{\infty}{\underset{k=1}{\sum}} \frac{(-1)^{k-1} (k\alpha) \lambda^{k\alpha-1}}{\Gamma(\alpha k +1)}$,
where $\lambda>0$ and $0<\alpha\leq 1$ (Supplementary \cref{sec:ML}). It interpolates between the exponential distribution, recovered at $\alpha=1$, and increasingly broad heavy-tailed distributions as $\alpha$ decreases. For $\alpha<1$, the Mittag-Leffler distribution has a power-law tail at large rates, $g(\lambda)\sim \lambda^{-(1+\alpha)}$ \cite{Mittag_Leffler_Asymptotics, gorenflo2010mittagleffler}, as well as a singularity at small rates where $g(\lambda)\sim\lambda^{\alpha-1}$. Thus, smaller values of $\alpha$ produce broader fluctuations and a higher probability of exceptionally small and large rates. We focus on this distribution because it naturally generates strong rate fluctuations while remaining controlled by a single parameter; and because its Laplace transform with respect to the jump rate, $\tilde{g}(s)=1/(1+s^\alpha)$ \cite{Laplace_ML}, admits a particularly simple form that will prove instrumental in the analysis below.

To solve for the first-passage time, we apply the recipe. Starting with step 1, we need the probability to be at $x$ after $n$ steps, in the presence of the absorbing boundary. This probability has the following integral form, $P(x,n|x_0) = 2\int_0^1\cos^n(p\pi) \sin(p\pi x)\sin(p\pi x_0) dp$ \cite{feller1968introduction}. Next, we proceed to step 2 and convert steps to time. Summing over $n$, we identify the Taylor series of the exponential, leading to $ P(x,t|x_0, \Lambda_t) = 2\int_0^1 \exp\left(-\Lambda_t(1-\cos p\pi)\right) \sin(p\pi x)\sin(p\pi x_0) dp$. Lastly, we advance to the final step and average over all cumulative rates. This gives 
\begin{equation} \label{Prob_x_t_half_line}
    P(x,t|x_0)  = 2 \int_0^1 \sin(p\pi x_0)\sin(p\pi x) \tilde{\Lambda}_t(1-\cos p\pi) dp,
\end{equation}
where $\tilde{\Lambda}_t(1-\cos\pi p)$ is the Laplace transform of the random variable $\Lambda_t$ evaluated at $(1-\cos\pi p)$. Further details are in   Supplementary~\cref{Section:Propagator_Half_Line}. Finally, summing over the non-absorbing sites, $x=1,2,3,...$, gives the survival probability (Supplementary \cref{Section:Survival_Probability_Half_Line}),
\begin{equation} \label{Survival_Prob_half_line}
     S(t|x_0) =  \int_0^1 \sin(p \pi x_0) \frac{\sin(p\pi )}{1-\cos(p\pi)}  \Tilde{\Lambda}_t(1-\cos(p\pi))   dp.
\end{equation}

\begin{figure}[t]
\centering
\includegraphics[width=0.485\textwidth]{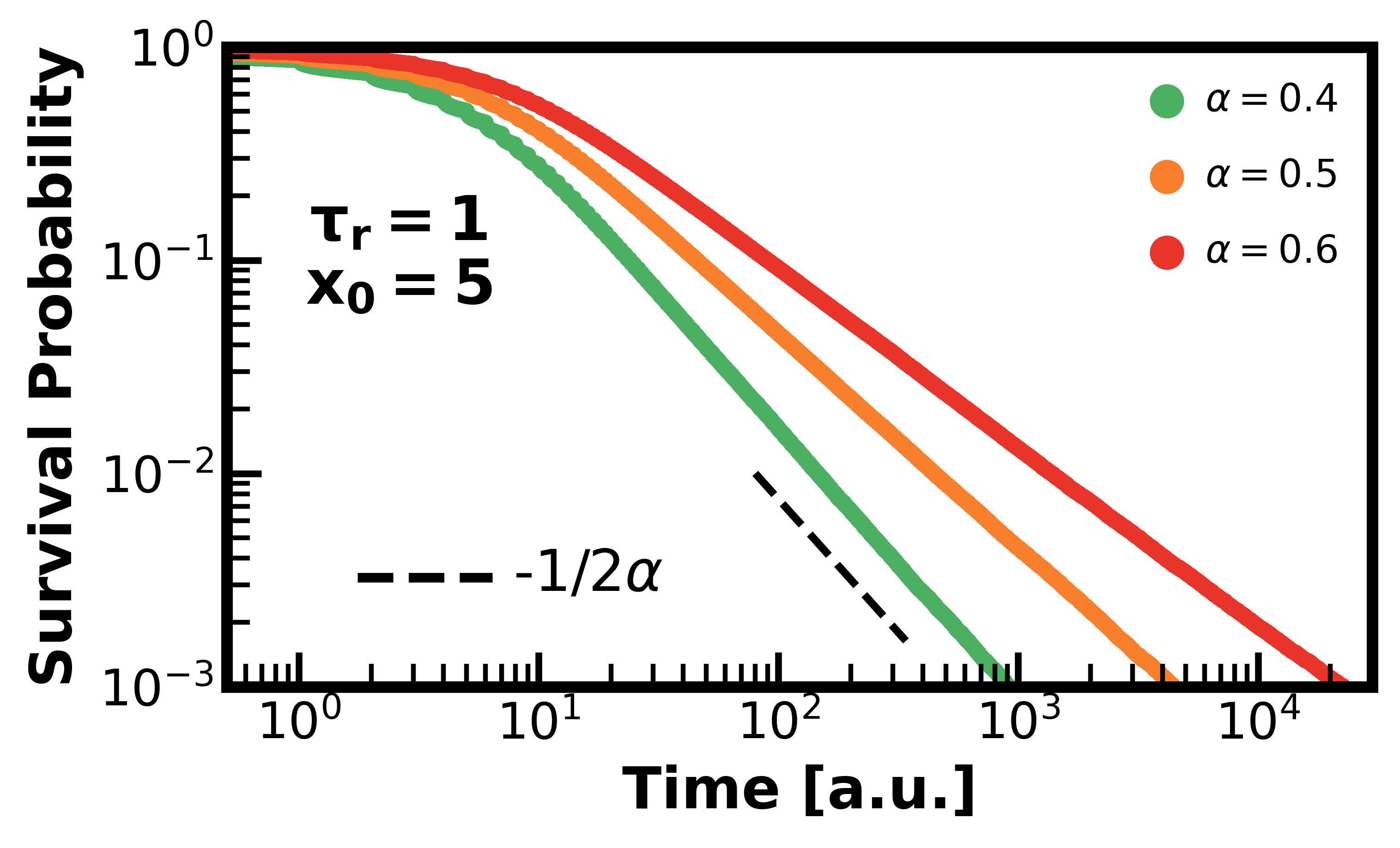}
 \caption{\textbf{Survival Probability}. Log-log plot of the survival probability as a function of time. Here, $\tau_r=1$ and $x_0=5$, and data come from simulations for $\alpha=0.4, 0.5, 0.6$ (green, orange, and red, respectively). Survival probability is based on one million walkers. A dashed line with slope $-1/2\alpha$, as predicted by \cref{Survival_Tail_Half-Line}, is shown parallel to the survival probability tail for $\alpha=0.4$.}
  \label{fig:Phase_Transition_survival}
\end{figure}

\Cref{Prob_x_t_half_line,Survival_Prob_half_line} apply to \textit{any} jump-rate dynamics. Specifically, when the jump rate is resampled every constant time interval $\tau_r$, from the Mittag-Leffler distribution, the Laplace transform of $\Lambda_t$ can be written as: $\tilde{\Lambda}_t(s) \equiv \braket{e^{-s\Lambda_t}} = \tilde{g}(s\cdot \tau_r)^{n_{\tau_r}}\cdot \tilde{g}(s\cdot \delta_t) = \left(\frac{1}{1+s^\alpha\tau_r^\alpha}\right)^{n_{\tau_r}} \frac{1}{1+s^\alpha\delta_t^\alpha}$, i.e., a product of Mittag-Leffler Laplace transforms. Plugging this into \cref{Survival_Prob_half_line} allows us to evaluate the tail of the survival probability 
(Supplementary  \cref{Section:Survival_Probability_Half_Line}), 
\begin{equation} \label{Survival_Tail_Half-Line}
    S(t|x_0) \simeq   \frac{x_0\sqrt{2} \tau_r^{(1-\alpha)/2\alpha}\Gamma\left(\frac{1}{2\alpha}\right)}{\alpha\pi}t^{-1/2\alpha} \sim t^{-1/2\alpha}.
\end{equation}
\cref{fig:Phase_Transition_survival}  shows this asymptotic behavior using simulations.  The prefactor and power-law exponent in \cref{Survival_Tail_Half-Line} are further verified in Supplementary  \cref{Section:Survival_Probability_Half_Line}. 

Observe that for $\alpha\geq \frac{1}{2}$, the power-law decay is very slow, yielding an infinite MFPT. For $\alpha< \frac{1}{2}$, the decay is sufficiently fast to result in a \textit{finite} MFPT. Thus, beyond the critical value of $\alpha_c=\frac{1}{2}$, there is a striking qualitative change in first-passage behavior: 
\begin{equation} \label{MFPT_Sum}
    \text{MFPT} =  \int_0^\infty S(t|x_0)dt =\left\{\begin{array}{cc}
        \text{finite} & \alpha<\frac{1}{2},\\
       \infty  & \text{else}. 
    \end{array}\right.
\end{equation}

\noindent Recently, a similar transition has been found for compounded random walks \cite{Compounded_FirstPassage,Compounded_Random_Walk}, where the number of steps taken within each time interval is assumed to be heavy-tailed. In the DSCTRW, large-rate fluctuations produce analogous bursts of activity, dynamically generating many consecutive steps in rapid succession.

Having established the MFPT phase transition, we now turn to study it for $\alpha<\frac{1}{2}$, i.e., when it is  finite. In this regime, the MFPT can be obtained by integrating the survival probability in \cref{Survival_Prob_half_line} with respect to time.  Doing so we get (Supplementary \cref{Section:MFPT_Half_Line})
\begin{equation} \label{MFPT}
    \text{MFPT}(\alpha,\tau_r, x_0) = 
        \int_0^1 dp I_1(\alpha, \tau_r,x_0,p)\cdot  I_2(\alpha,\tau_r,p), 
\end{equation}
where $I_1(\alpha, \tau_r,x_0,p) \equiv \sin(p \pi x_0) \frac{\sin(p\pi )}{1-\cos(p\pi)}  \frac{1+\left[\left(1-\cos p\pi\right)\tau_r\right]^\alpha}{\left[\left(1-\cos p\pi\right)\tau_r\right]^\alpha}$ and $I_2(\alpha,\tau_r, p) \equiv \int_{0}^{\tau_r}\frac{dz}{1+[1-\cos\left(p\pi\right)]^\alpha z^\alpha}$. From \cref{MFPT} we can uncover how the MFPT scales with  the distance from the origin. For asymptotically large $x_0$, we find (Supplementary  \cref{Sec:MFPT_Scaling_x0}), 
\begin{equation} \label{eq:MFPT_x0}
     \text{MFPT}(\alpha<0.5, \tau_r, x_0) \simeq \
         C_1 \cdot x_0^{2\alpha},  
\end{equation}
with $C_1=\frac{2^\alpha \tau_r^{1-\alpha}\Gamma(1-2\alpha)\sin(\pi\alpha)}{\pi\alpha}$.  Notably,  the MFPT grows sublinearly with $x_0$. We showcase this in \cref{fig:Phase_Transition}(a), and further verify the prefactor and power-law exponent in Supplementary \cref{Sec:MFPT_Scaling_x0}.

\begin{figure*}[t]
\centering
\includegraphics[width=\textwidth]{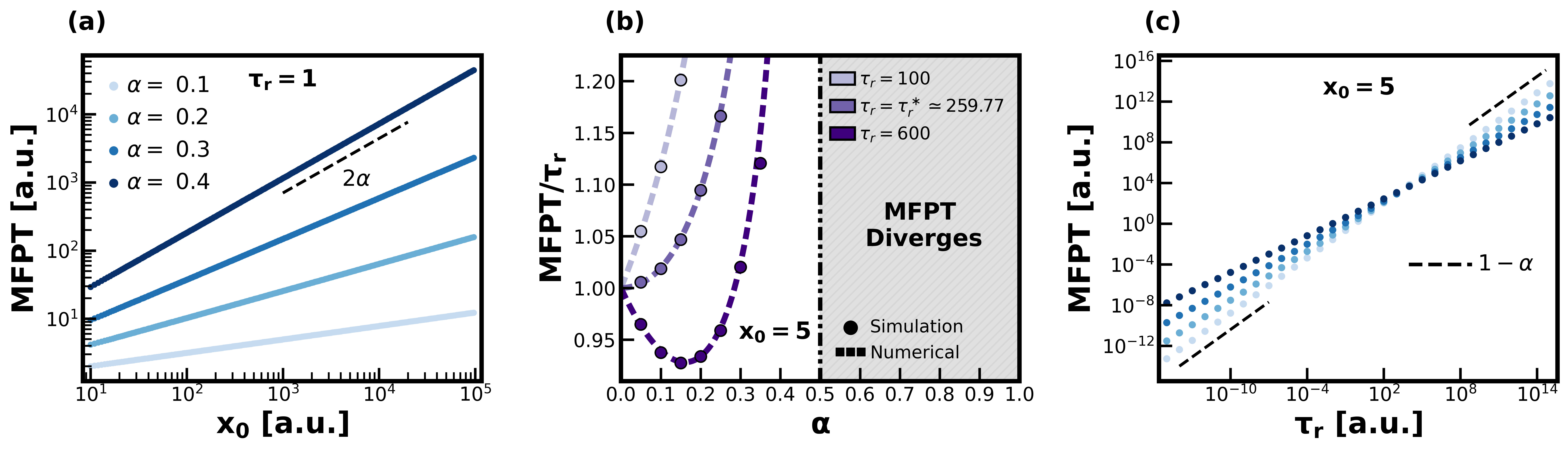}
  \caption{\textbf{MFPT Phase Transition}. \textbf{a.} Log-log plot of the MFPT as a function of $x_0$ for $\tau_r=1$. Numerical evaluations of \cref{MFPT} for four different values of $\alpha$ are shown in blue, from $\alpha=0.1$ (lightest blue) to $\alpha=0.4$ (darkest blue); the dashed line shows the $2\alpha$ slope attained at  $x_0\gg 1$ for $\alpha=0.4$. \textbf{b.} MFPT/$\tau_r$ as a function of $\alpha$ for three different values of $\tau_r$: small ($100$), critical ($\simeq 259.77$) and large ($600$), from lightest  to darkest purple. Data coming from simulations is in circles, and dashed lines come from numerical evaluation of equation \cref{MFPT}. Here, $x_0=5$ and simulations are based on one million walkers. For $\alpha\geq \frac{1}{2}$ the MFPT diverges, as indicated by the gray shaded area. \textbf{c.} Log-log plot of the MFPT as a function of $\tau_r$ for $x_0=5$, for the same values of $\alpha$ as in (a).  Numerical evaluations are in blue, with dashed lines showing the slope of $1-\alpha$, in the case $\alpha=0.1$, for small and large $\tau_r$.}
  \label{fig:Phase_Transition}
\end{figure*}

In \cref{fig:Phase_Transition}(b) we plot the MFPT, normalized by $\tau_r$, as a function of $\alpha$, and verify \cref{MFPT} against simulation data. As predicted,  when $\alpha\shortrightarrow \frac{1}{2}$ the MFPT diverges. \cref{MFPT} allows us to identify the origin of this divergence and characterize its scaling. Specifically, for $p \ll 1$ the integrand scales as $  \sim \frac{1}{p^{2\alpha}}$, thus in the limit $\alpha\shortrightarrow \frac{1}{2}$  we have MFPT $\sim \frac{1}{1-2\alpha}$. Fits illustrating this scaling, along with the full derivation, are provided in Supplementary \cref{Section:MFPT_Half_Line_Alpha_limits}.

\cref{fig:Phase_Transition}(b) exhibits rich behavior of the MFPT. First, as $\alpha\shortrightarrow 0$, the MFPT converges to $\tau_r$ regardless of its value. This is not a coincidence.  In this limit, the Mittag-Leffler distribution becomes increasingly concentrated at two extremes: rate zero and rate infinity. The probability of sampling either converges to half. Thus, sampling a rate is analogous to flipping a coin: heads for rate zero; tails for rate infinity.  When we flip tails (rate infinity), the walker reaches the target in zero time. This occurs on average after two flips. As we first flip at $t=0$, and then again every $\tau_r$,  the second flip is at $t=\tau_r$. Therefore, the MFPT is  $\tau_r$. A detailed derivation of this result is provided in Supplementary \cref{Section:MFPT_Half_Line_Alpha_limits}, along with additional numerical verification. 

Next,  we note two distinct behaviors of the MFPT. For the smallest $\tau_r$ (lightest purple) the MFPT monotonically increases. For the largest $\tau_r$ (darkest purple) the MFPT decreases at $\alpha=0$ until it reaches a minimum at some $\alpha^\ast\neq 0$, and then increases until diverging as $\alpha\shortrightarrow \frac{1}{2}$. While there is no simple formula for $\alpha^*$, we find that the transition between the two behaviors occurs at a critical relaxation time: $\tau_r^\ast= \exp\left(\ln2 +2H_{x_0}-\frac{1}{x_0}+\frac{1}{2}\right) \propto x_0^2$, where $H_{x_0} = \overset{x_0}{\underset{k=1}{\sum}}\frac{1}{k}\sim \ln(x_0) $ is the $x_0$-th Harmonic number \cite{HarmonicNumber3} (see Supplementary  \cref{Section:MFPT_Half_Line_Minimum}). 

We now investigate how  the relaxation time affects the MFPT. Interestingly, for a given $\alpha<0.5$,  the MFPT exhibits the same power-law in the two extremes of  asymptotically short and long relaxation times (Supplementary \cref{Section:MFPT_Half_Line_Tau_r_limits}),
\begin{equation} \label{eq:MFPT_Asym}
    \text{MFPT}(\alpha<0.5, \tau_r, x_0) \simeq \left\{ \begin{array}{cc}
         \frac{C_0}{1-\alpha} \cdot \tau_r^{1-\alpha}& \tau_r\gg 1 , \\[9pt]
         C_0  \cdot \tau_r^{1-\alpha}&\tau_r\ll1, 
    \end{array}\right. 
\end{equation}

where $C_0=\int_0^1 \sin(p \pi x_0) \frac{\sin(p\pi )}{\left[1-\cos(p\pi)\right]^{1+\alpha}}dp  $. \cref{fig:Phase_Transition}(c) shows that the MFPT scales in agreement with \cref{eq:MFPT_Asym} {(prefactor verification is provided in Supplementary \cref{Section:MFPT_Half_Line_Tau_r_limits})}.
When $\tau_r$ increases (decreases) the MFPT increases (decreases) with it. We can understand this intuitively. When the rate changes rarely, we approach the classical case of a constant rate CTRW, where the MFPT diverges. In the other extreme, frequent rate changes increase the chance of sampling extremely large rates from the heavy-tail of the Mittag-Leffler distribution, enabling bursts of many steps within a single interval $\tau_r$, thus accelerating first-passage.

\label{Section:Conclusions}
\textit{Conclusions and outlook.---}In this letter, we investigated how fluctuations in the jump rate affect the first-passage properties of a  simple symmetric one-dimensional random walk. Within the DSCTRW framework, we showed that sufficiently broad rate fluctuations fundamentally alter first passage: the MFPT undergoes a transition at $\alpha_c=1/2$, becoming finite for $\alpha<1/2$ despite diverging for the ordinary random walk. The transition originates from rare, exceptionally large jump rates, which generate bursts of rapid motion that regularize the long-time first-passage tail and render its mean finite.

Importantly, this transition extends well beyond the specific model considered here. It does not rely on the Mittag-Leffler rate distribution used to prove \cref{Survival_Tail_Half-Line}. Any distribution with  large-rate asymptotics, $g(\lambda)\sim\lambda^{-1-\alpha}$, exhibits the same transition at $\alpha_c=1/2$, irrespective of its behavior at small rates (see \cite{appendixA}). Moreover, the transition is expected to persist beyond piecewise constant stochastic rate processes, as long as they have a finite relaxation time. Finally, the same mechanism should carry over to continuous space, suggesting an analogous first-passage transition for diffusion, e.g., in the framework of diffusing diffusivity \cite{DiffusingDiffusivitySlater,DD_Superstatistics} or switching diffusion \cite{gueneau2025large}. Together, these observations point to a general mechanism by which sufficiently strong diffusivity fluctuations can regularize otherwise divergent first-passage times.

\textit{Acknowledgments.---}This project has received funding from the European Research Council (ERC) under the European Union’s Horizon 2020 research and innovation program (Grant Agreement No. 947731). B.L is grateful for the support of the Colton family, for their generous Colton Family Foundation scholarship, and for the Planning and Budgeting Committee of the Council for Higher Education for their scholarship for Outstanding Women Doctoral Students in the High-Tech.

\clearpage

\section*{End Matter} 

\paragraph{Appendix A: Universality of the mean FPT transition.---}We show that for any heavy-tailed distribution with large-rate power-law asymptotics $\sim \lambda^{-1-\alpha}$ the MFPT becomes finite for $\alpha<\frac{1}{2}$.  
We start with the full expression of the survival probability in \cref{Survival_Prob_half_line}. Recall that $\Lambda_t$ is expressed as a sum of the rates (multiplied by the duration of each rate), thus $\tilde{\Lambda}_t$ is the product of the Laplace transform of the rate distribution. Since we are interested in large times, $t\gg \tau_r$, we have $n_{\tau_r}= \left\lfloor \frac{t}{\tau_r}\right\rfloor \simeq \frac{t}{\tau_r}\gg 1$, where  the Laplace transform $\tilde{\Lambda}_t$ in the integrand rapidly decays from a maximal value of unity as the integral variable $p$ is increased from zero. This means that the main contribution to the integral comes from small $p$. Expanding the integrand to leading order, the survival probability becomes (Supplementary \cref{Section:Survival_Probability_Half_Line})
\begin{equation}  
S(t|x_0) \simeq 2x_0 \int_0^1   \left[\tilde{g}\left(\frac{(p\pi)^2\tau_r}{2}\right)\right]^{n_{\tau_r}} \tilde{g}\left(\frac{(p\pi)^2\delta_t}{2}\right)dp.
\end{equation}

\noindent From the Tauberian theorem \cite{Tauberian_Fourier, klafter2011first}, we know that if the distribution has a tail behavior $\sim\lambda^{-1-\alpha}$ then its Laplace transform for $s\ll 1$ is $\tilde{g}(s)\simeq 1-C s^\alpha\simeq \exp(-Cs^\alpha)$, where $C$ is a positive prefactor. Here we require $s\equiv \frac{(p\pi)^2\tau_r}{2} \ll 1$, and as $\delta_t\leq \tau_r$ it immediately follows that $\frac{(p\pi)^2\delta_t}{2} \ll 1$.

\noindent Thus,
\begin{equation}
    S(t|x_0) \simeq  2x_0\int_0^1   \exp\left(-C\frac{(p\pi)^{2\alpha}}{2^\alpha} \left(n_{\tau_r}\tau_r^{\alpha} +\delta_t^\alpha\right)\right) dp.
\end{equation}

\noindent Define $A\equiv \frac{Cn_{\tau_r}\tau_r^{\alpha-1}\pi^{2\alpha}}{2^\alpha} $ and perform an integral variable change $z\equiv p^{2\alpha}A$,
\begin{equation}
    S(t|x_0) \simeq \frac{x_0}{\alpha} A^{-1/2\alpha}  \int_0^{A}    \exp\left\{-z\left(1+\frac{\delta_t^\alpha}{t\tau_r^{\alpha-1}}\right)\right\} z^{\frac{1}{2\alpha}-1} dz.
\end{equation}

\begin{figure}[t]
\centering
\includegraphics[width=0.45\textwidth]{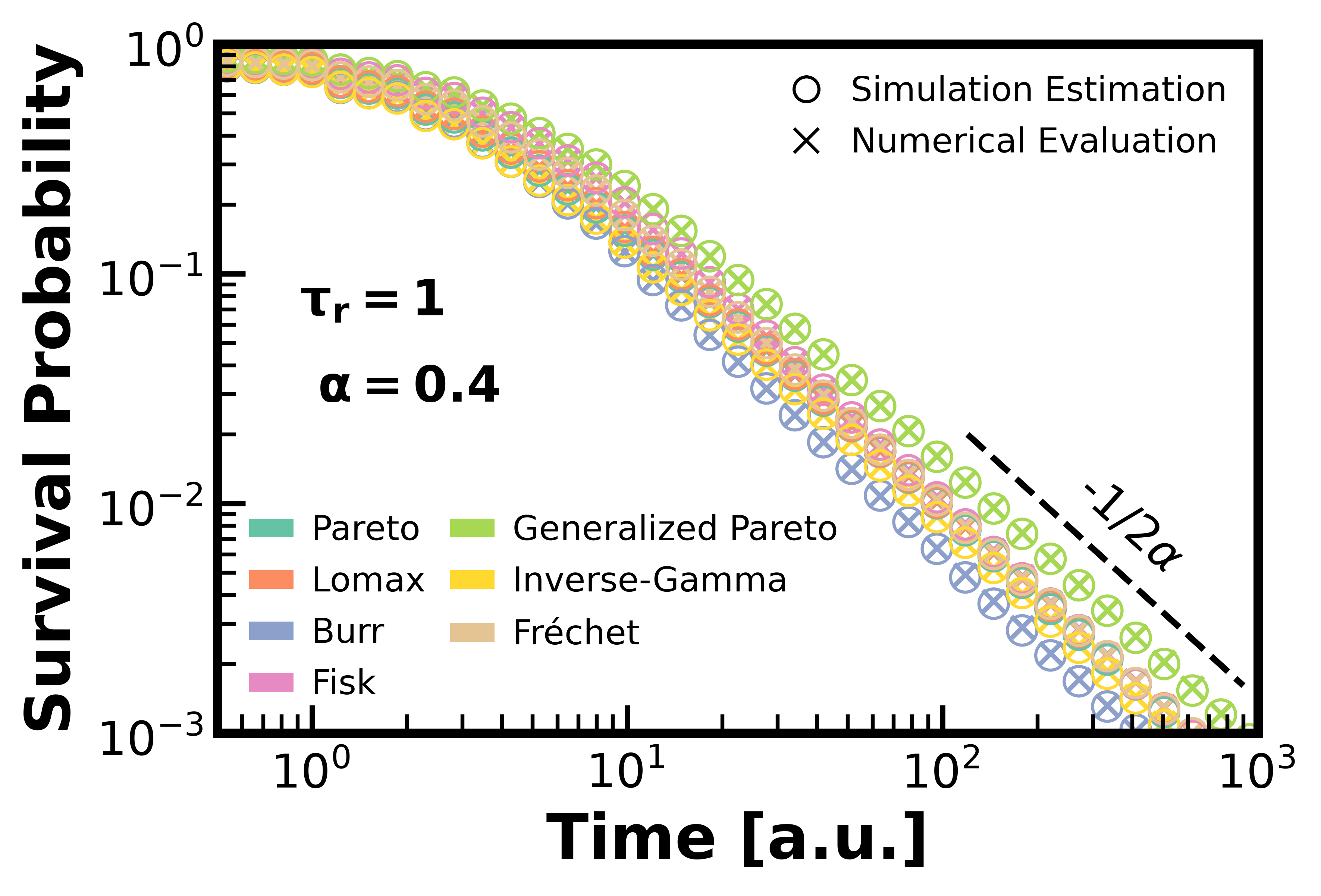}
  \caption{\textbf{Survival probability as a function of time.}  The survival probability as a function of time for seven different jump rate distributions, all with the same heavy-tail asymptotics $\sim \lambda^{-\alpha-1}$. Details of the distributions used  are given in \cref{tab:distributions}. Curves are based on data from simulations of 1 million walkers starting at $x_0=5$ for $\tau_r=1$ and $\alpha=0.4$. A dashed line indicates a power-law of (-1/2$\alpha$). }
  \label{fig:universality_heavy_tail}
\end{figure}

\noindent Since we are interested in large time asymptotics, we have $A\gg 1$. We thus take the upper integration boundary to infinity, to obtain
\begin{equation}
    \begin{split} S(t|x_0) &\simeq \frac{x_0}{\alpha} A^{-1/2\alpha}  \int_0^{\infty}    \exp\left\{-z\left(1+\frac{\delta_t^\alpha}{t\tau_r^{\alpha-1}}\right)\right\} z^{\frac{1}{2\alpha}-1}   dz= \\
&=\frac{x_0}{\alpha} A^{-1/2\alpha}  \Gamma\left(\frac{1}{2\alpha}\right) \frac{1}{\left(1+\frac{\delta_t^\alpha}{t\tau_r^{\alpha-1}}\right)^{1/2\alpha}},\end{split}
\end{equation}
where in the last equality we identified the Gamma function. Note that since $\delta_t\leq \tau_r$ then $\frac{\delta_t^\alpha}{t\tau_r^{\alpha-1}} \leq \frac{\tau_r^\alpha}{t\tau_r^{\alpha-1}} = \frac{\tau_r}{t}\ll 1$. Plugging in $A$, we find that the survival probability is approximately given by 
\begin{equation}
    S(t|x_0) \simeq   \frac{  x_0\sqrt{2} \tau_r^{(1-\alpha)/2\alpha}}{C^{1/2\alpha}\alpha\pi}\Gamma\left(\frac{1}{2\alpha}\right)t^{-1/2\alpha} \sim  t^{-1/2\alpha}.
\end{equation}
The scaling is identical to the one obtained with the Mittag-Leffler distribution. The difference is in the prefactor, where in the general case we have the constant $C$ from the Laplace transform small $s$ expansion.

In \cref{fig:universality_heavy_tail} we plot the survival probability as a function of time, for seven different distributions, all with the same heavy-tail scaling $\sim \lambda^{-\alpha-1}$ (\cref{tab:distributions}). Although these distributions differ in their bulk, and in their behavior at low rates, the tails of the corresponding survival probabilities collapse onto a single power law with exponent $-1/2\alpha$.

\begin{table*}[!t]
\centering
\renewcommand{\arraystretch}{2.7}

\begin{tabularx}{\textwidth}{
    >{\raggedright\arraybackslash}p{2.7cm}
    >{\centering\arraybackslash}X
    >{\centering\arraybackslash}p{2cm}
    >{\centering\arraybackslash}p{2.2cm}
}
\toprule
\textbf{Distribution}
&
\textbf{Probability density function}
&
\textbf{Scale parameter}
&
\textbf{Shape parameter(s)}
\\
\midrule

Pareto
&
$\displaystyle
f(\lambda)=\frac{\alpha x_m^\alpha}{\lambda^{\alpha+1}},
\qquad \lambda\geq x_m
$
&
$x_m=1$
&
$\alpha>0.4$
\\

Lomax
&
$\displaystyle
f(\lambda)=\frac{\alpha}{\sigma}
\left(1+\frac{\lambda}{\sigma}\right)^{-(\alpha+1)},
\qquad \lambda\geq0
$
&
$\sigma=1$
&
$\alpha=0.4$
\\

Burr (Type III)
&
$\displaystyle
f(\lambda)=\frac{cd}{\sigma}\left(\frac{\lambda}{\sigma}\right)^{-c-1}
\left[1+\left(\frac{\lambda}{\sigma}\right)^{-c}\right]^{-d-1},
\qquad \lambda>0
$
&
$\sigma=1$
&
$c=0.4,d=1.5$
\\

Fisk 
&
$\displaystyle
f(\lambda)=\frac{\beta}{\sigma}
\left(\frac{\lambda}{\sigma}\right)^{\beta-1}
\left[1+\left(\frac{\lambda}{\sigma}\right)^\beta\right]^{-2},
\qquad \lambda>0
$
&
$\sigma=1$
&
$\beta=0.4$
\\

Generalized Pareto
&
$\displaystyle
f(\lambda)=\frac{1}{\sigma}
\left(1+\xi\frac{\lambda}{\sigma}\right)^{-1/\xi-1}
$
&
$\sigma=1$
&
$\xi=1/0.4$
\\

Inverse Gamma
&
$\displaystyle
f(\lambda)=
\frac{\beta^\alpha}{\Gamma(\alpha)}
\lambda^{-(\alpha+1)}
\exp\left(-\frac{\beta}{\lambda}\right),
\qquad \lambda>0
$
&
$\beta=1$
&
$\alpha=0.4$
\\

Fréchet
&
$\displaystyle
f(\lambda)=\frac{\alpha}{s}
\left(\frac{\lambda}{s}\right)^{-1-\alpha}
\exp\left[-\left(\frac{\lambda}{s}\right)^{-\alpha}\right],
\qquad \lambda>0
$
&
$s=1$
&
$\alpha=0.4$
\\

\bottomrule
\end{tabularx}
\caption{\textbf{Probability density functions and their parameters.} The jump rate probability density functions used in \cref{fig:universality_heavy_tail}, along with the chosen scale and shape parameters. } \label{tab:distributions}
\end{table*}

\clearpage
\onecolumngrid


\newpage
\onecolumngrid

\begin{center}
    {\Large\bfseries Supporting Information}\\[8pt]
    {\large First Passage in the Presence of Jump Rate Fluctuations:\\[4pt]
    A Transition from Infinite to Finite Mean}\\[12pt]
    Bara Levit$^1$ \quad and \quad Shlomi Reuveni$^1$\\[4pt]
    \textit{$^1$School of Chemistry, The Center for Computational Molecular and Materials Science,}\\
    \textit{The Center for Physics and Chemistry of Living Systems,}\\
    \textit{Tel Aviv University, Tel Aviv, Israel}
\end{center}

\setcounter{section}{0}
\setcounter{equation}{0}
\setcounter{figure}{0}
\setcounter{table}{0}

\renewcommand{\thesection}{S\arabic{section}}
\renewcommand{\theequation}{S\arabic{equation}}
\renewcommand{\thefigure}{S\arabic{figure}}
\renewcommand{\thetable}{S\arabic{table}}

\section{Piecewise Constant Stochastic Rate} \label{sec:Piecewise_Stochastic_Rate}
We focus on the scheme where the rate changes every constant time $\tau_r$ --- the relaxation time. In this case the cumulative rate $\Lambda_t$ can be written as 
\begin{equation}
    \Lambda_t\equiv \int_0^t\lambda_{t^\prime} dt^\prime = \sum_{i=0}^{n_{\tau_r}-1}\lambda_{i\cdot \tau_r}\tau_r + \lambda_{n_{\tau_r}\cdot \tau_r}\delta_t,
\end{equation}
where $n_{\tau_r} = \left\lfloor\frac{t}{\tau_r}\right\rfloor$, $\delta_t = t-n_{\tau_r}\tau_r$ and $\lambda_{i\cdot \tau_r}$ are independent and identically distributed (i.i.d) random variables sampled from some distribution $g(\lambda)$.

The Laplace transform of a sum of i.i.d random variables is the product of the Laplace transforms of the individual random variable $\tilde{g}(s)=\int_0^\infty e^{-s\lambda}g(\lambda)d\lambda$. This gives

\begin{equation} \label{Laplace_Transform_Product}
    \tilde{\Lambda}_t(s) \equiv \braket{e^{-s\Lambda_t}}= [\tilde{g}(s\cdot \tau_r)]^{n_{\tau_r}} \tilde{g}(s\cdot \delta_t).
\end{equation}

\section{Mittag-Leffler Distribution} \label{sec:ML}

The pdf of the Mittag-Leffler  distribution is given by \cite{ML_Distribution1, gorenflo2010mittagleffler}: 
\begin{equation}
g(\lambda)=\sum_{k=1}^\infty \frac{(-1)^{k-1} (k\alpha) \lambda^{k\alpha-1}}{\Gamma(\alpha k +1)},
\label{eq:ML}
\end{equation}
\noindent
where $\lambda>0$ and $0<\alpha\leq 1$. For $\alpha=1$ the Mittag-Leffler distribution reduces to the exponential distribution \cite{ML_Distribution1}. For $\alpha<1$, the Mittag-Leffler distribution has a heavy tail and a diverging mean. The heavy-tail  power-law  of the Mittag-Leffler distribution is given by \cite{Mittag_Leffler_Asymptotics, gorenflo2010mittagleffler} 
\begin{equation} 
g(\lambda)\sim \frac{\Gamma(\alpha+1)\sin(\alpha\pi)}{\pi}\lambda^{-\alpha-1} \qquad \lambda \gg 1.
\label{eq:ML-large-lambda}
\end{equation}
\noindent The smaller $\alpha$ is, the `heavier' the power-law.  Similarly, there exists a power-law for small values as well \cite{Mittag_Leffler_Asymptotics,gorenflo2010mittagleffler},
\begin{equation}
g(\lambda)\sim \frac{\alpha}{\Gamma(1+\alpha)}\lambda^{\alpha-1} = \frac{1}{\Gamma(\alpha)} \lambda^{\alpha-1} \qquad \lambda \ll 1.
\label{eq:ML-small-lambda}
\end{equation}

\noindent The Laplace transform of a Mittag-Leffler random variable, $\lambda$, is given by  \cite{Laplace_ML}
\begin{equation} \label{ML-Laplace_Transform}
    \tilde{g}(s) = \frac{1}{1+s^\alpha}.
\end{equation}

\begin{figure}[H]
\centering
\includegraphics[width=0.45\textwidth]{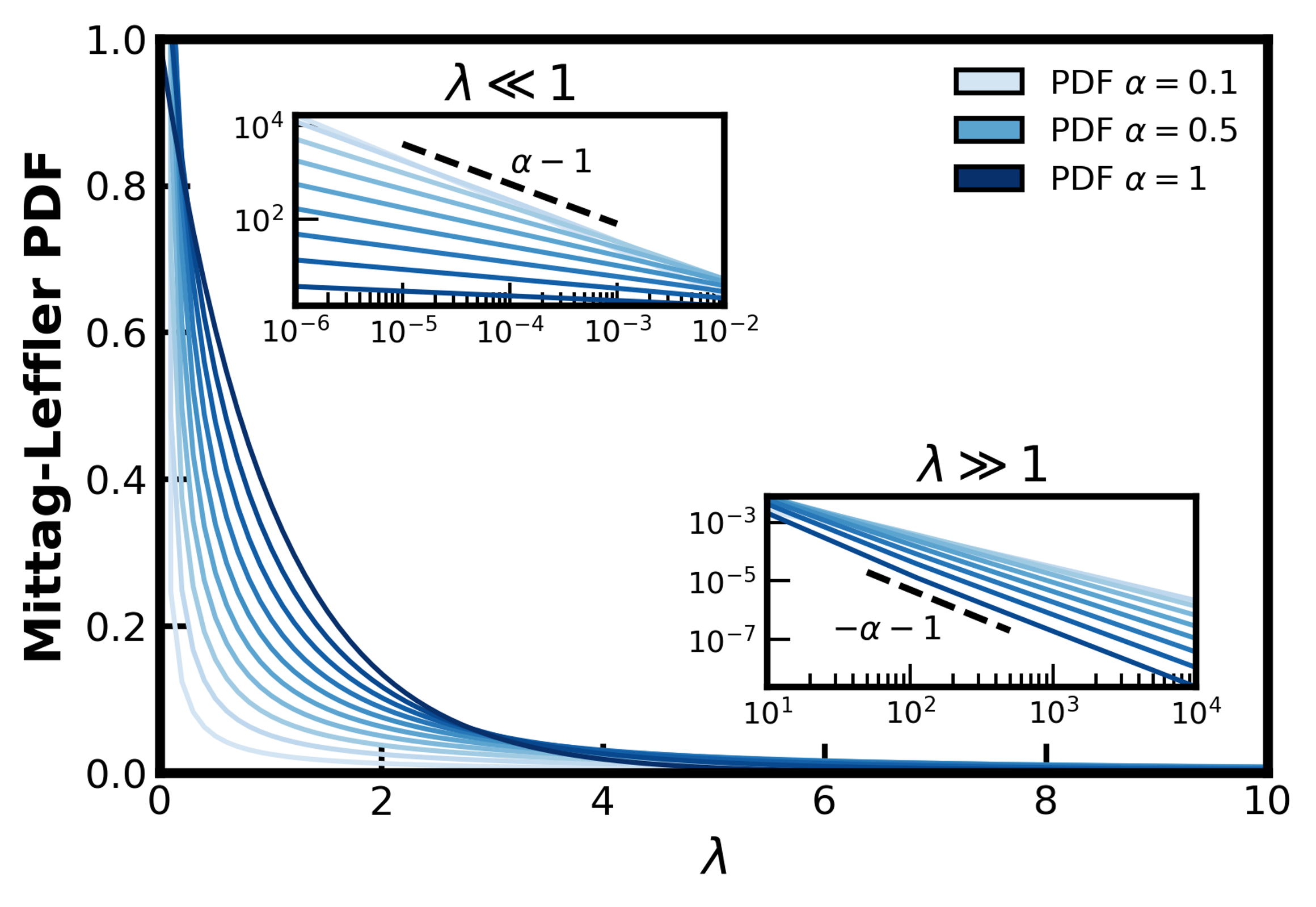}
  \caption{\textbf{Mittag-Leffler Distribution.}  The Mittag-Leffler pdf for different values of $\alpha$ according to \cref{eq:ML}. Darkest shade  corresponds to $\alpha=1$ and as the shade gets lighter $\alpha$ decreases. The inset in the top (bottom) left (right) shows the power law for small (large) $\lambda$. Dashed line is provided as a guide to the slope.  }
  \label{fig:ML}
\end{figure}

In \cref{fig:ML}, we show the pdf of the Mittag-Leffler distribution for different values of $\alpha$. The darkest blue corresponds to $\alpha=1$, and as the blue gets lighter the values of $\alpha$ get smaller. In the insets, we show the power-law behavior for small and large $\lambda$, \cref{eq:ML-large-lambda,eq:ML-small-lambda}.

\section{First Passage on the Half-Line} \label{section:Half-Line_SI}
\subsection{Propagator on the Half-Line (Derivation of \cref{Prob_x_t_half_line})} \label{Section:Propagator_Half_Line}

The simple symmetric random walk in 1D, starting at $x_0$, with an absorbing boundary at $x=0$, also known as the simple symmetric random walk on the half-line, has a known propagator. The more acquainted one has combinatorial terms (can be obtained using the image method, for instance) \cite{Prop_1D_Common_Combi}. Another form exists, obtained from the propagator in the interval $[0,L]$, taking the limit $L\rightarrow \infty$  
\cite{feller1968introduction}, or alternatively by utilizing the image method on the inverse Fourier transform of $\phi^n(k)$ (the characteristic function of a single step raised to the power of $n$). The  expression is  
\begin{equation} \label{half-line-propogator}
    P(x,n|x_0) = 2\int_0^1\cos^n(p\pi) \sin(p\pi x)\sin(p\pi x_0) dp.
\end{equation}

We now proceed to the second step: translating the steps into time. We do this by multiplying \cref{half-line-propogator} with the Poisson distribution with mean $\Lambda_t$ (the cumulative rate) and then summing over possible steps. This gives the probability to be at $x$ at time $t$, given a specific cumulative jump rate  $\Lambda_t$, 
\begin{equation}
\begin{split}\label{SS_Half-line_Step2}
    P(x,t|x_0, \Lambda_t)&= \sum_{n=0}^\infty P(x,n|x_0)\chi_n(t) = \\
    &= 2\int_0^1 \sin(p\pi x_0)\sin(p\pi x) \exp\left\{-\Lambda_t\left(1-\cos\pi p \right)\right\} dp.
\end{split}
\end{equation}
The last equality was obtained by identifying the Taylor expansion series of the exponential function. Note that in the expression for the propagator of the simple symmetric random walk \cref{half-line-propogator}, the cosine is raised to the power of $n$. This property is helpful as we can gather all terms raised to the power of $n$ (in the propagator and in the Poisson distribution), and utilize the sum over $n$ to identify the Taylor expansion of the exponential function. 

Lastly to obtain the full probability, we need to average with respect to the cumulative rate $\Lambda_t$. Doing so, we obtain Eq.~\eqref{Prob_x_t_half_line} in the main text
\begin{equation}\label{half-line_Double_Propagator}
P(x,t|x_0) = \left\langle  P(x,t|x_0, \Lambda_t) \right\rangle_{\Lambda_t}=2\int_0^1 dp \sin(p\pi x_0)\sin(p\pi x)   \Tilde{\Lambda}_t(1-\cos\pi p),
\end{equation}
where $\Tilde{\Lambda}_t(1-\cos\pi p)$ is the Laplace transform of the random variable $\Lambda_t$, evaluated at $(1-\cos\pi p)$.

\subsection{Survival Probability on the Half-Line (Derivation of \cref{Survival_Prob_half_line} and \cref{Survival_Tail_Half-Line})} \label{Section:Survival_Probability_Half_Line}

To find the survival probability one sums the propagator in \cref{half-line_Double_Propagator} over all possible lattice sites. 
\begin{equation}
    \begin{split}
S(t|x_0)&=\sum_{x=1}^{\infty}P(x,t) = \sum_{x=1}^{\infty} 2 \int_0^1 \sin(p\pi x_0)\sin(p\pi x) \tilde{\Lambda}_t(1-\cos(p\pi)) dp =    \\
     &=2  \int_0^1dp \sin(p\pi x_0) \tilde{\Lambda}_t(1-\cos(p\pi))   \sum_{x=1}^{\infty} \sin(p\pi x). \end{split}
\end{equation}
We now focus on the sum with respect to $x$ over the sine expression,
\begin{equation}
    \begin{split}
     \sum_{x=1}^{\infty}\sin(p\pi x)&= \sum_{x=1}^{\infty} \left\{e^{ip\pi x} - e^{-ip\pi x}\right\} =\frac{1}{2i}  \left\{\frac{1}{1-e^{ip\pi }} -1- \frac{1}{1-e^{-ip\pi }}+1\right\}=\frac{1}{2i}  \left\{\frac{1-e^{-ip\pi } - 1+e^{ip\pi } }{1-e^{ip\pi } + 1-e^{-ip\pi }}\right\}= \\
     &= \frac{1}{2i}  \left\{\frac{e^{ip\pi }-e^{-ip\pi }}{2-(e^{ip\pi x} + e^{-ip\pi })}\right\}=\frac{1}{2i}  \left\{\frac{2i\sin(p\pi )}{2-2\cos(p\pi )}\right\}=\frac{\sin(p\pi )}{2-2\cos(p\pi )}.
\end{split}
\end{equation}
Plugging this expression back to the survival probability we get 
\begin{equation}
    \begin{split}
    S(t|x_0)&=2  \int_0^1dp \sin(p\pi x_0) \tilde{\Lambda}_t(1-\cos(p\pi)) \frac{\sin(p\pi )}{2-2\cos(p\pi )}\\ 
     & =\int_0^1dp \sin(p\pi x_0) \frac{\sin(p\pi )}{1-\cos(p\pi )} \tilde{\Lambda}_t(1-\cos(p\pi)).
\end{split}
\end{equation}

\noindent Thus, as expressed in Eq.~\eqref{Survival_Prob_half_line} in the main text, the survival probability for the doubly stochastic walk in the half-line is
\begin{equation} \label{Final_Expression_Survival_Probability_Half-Line}
     S(t|x_0) =  \int_0^1 \sin(p \pi x_0) \frac{\sin(p\pi )}{1-\cos(p\pi)}  \Tilde{\Lambda}_t(1-\cos(p\pi))   dp.
\end{equation}

\noindent This result is general, whether the rate is changed every constant time $\tau_r$ or otherwise, and for any jump rate distribution.

We adopt the scheme of resampling the rate every constant time interval $\tau_r$. Utilizing \cref{Laplace_Transform_Product} and \cref{ML-Laplace_Transform} we obtain
\begin{equation}
     S(t|x_0) \simeq  \int_0^1 \sin(p \pi x_0) \frac{\sin(p\pi )}{1-\cos(p\pi)} \left[\tilde{g}((1-\cos(p\pi))\tau_r)\right]^{n_{\tau_r}} \tilde{g}((1-\cos(p\pi))\delta_t) dp.\end{equation}
\noindent Since we are interested in large times, $t\gg \tau_r$, we have $n_{\tau_r}= \left\lfloor \frac{t}{\tau_r}\right\rfloor \simeq \frac{t}{\tau_r}\gg 1$, where  the Laplace transform $\tilde{\Lambda}_t$ in the integrand rapidly decays from a maximal value of unity as the integral variable $p$ is increased from zero \cite{fellerintroduction_vol2}. This means that the main contribution to the integral comes from small $p$. Expanding the integrand to leading order gives
\begin{equation} \label{survival_probability_half_line_large_t}
     S(t|x_0) \simeq   2x_0 \int_0^1   \left[\tilde{g}\left(\frac{(p\pi)^2\tau_r}{2}\right)\right]^{n_{\tau_r}} \tilde{g}\left(\frac{(p\pi)^2\delta_t}{2}\right) dp .
\end{equation}

\subsubsection{Mittag leffler} \label{Survival_Probability_Mittag_Leffler}
We now consider the case where the rates are sampled from the Mittag-Leffler distribution. We will evaluate the survival probability for $t\gg\tau_r$. To do so we need to plug \cref{ML-Laplace_Transform} in  \cref{survival_probability_half_line_large_t}, for $s=\frac{(p\pi)^2\tau_r}{2}$ and $s=\frac{(p\pi)^2\delta_t}{2}$. We obtain 
\begin{equation}
    S(t|x_0) \simeq  2x_0 \int_0^1   \left[\frac{1}{1+\left(\frac{(p\pi)^2\tau_r}{2}\right)^\alpha}\right]^{n_{\tau_r}} \frac{1}{1+\left(\frac{(p\pi)^2\delta_t}{2}\right)^\alpha} dp = 2x_0 \int_0^1   \exp\left\{-n_{\tau_r} \ln\left[1+\frac{(p\pi)^{2\alpha}}{2^\alpha} \tau_r^\alpha \right]\right\} \frac{1}{1+\left(\frac{(p\pi)^2\delta_t}{2}\right)^\alpha} dp ,  
\end{equation}
where in the last equality we rewrote the fraction as an exponent. 

For $p\ll 1$ we can use the Taylor expansion for the logarithmic function, $\ln(1+x)\simeq x$, as well as the second term $\frac{1}{1+x}\simeq 1-x\simeq \exp(-x)$. 
\begin{equation}S(t|x_0) \simeq 2x_0\int_0^1 dp \exp\left\{-\frac{(p\pi)^{2\alpha}}{2^\alpha}\left(n_{\tau_r} \tau_r^\alpha+\delta_t\right)\right\}. 
\end{equation}
Denote $A\equiv \frac{t\tau_r^{\alpha-1}\pi^{2\alpha}}{2^\alpha}$ and perform an integral variable change $z\equiv p^{2\alpha}A$,
\begin{equation}S(t|x_0) \simeq2x_0\int_0^{A} \frac{dz}{2\alpha } z^{\frac{1-2\alpha}{2\alpha}} A^{-1/2\alpha} \exp\left\{-z\left(1+\frac{\delta_t^\alpha}{t\tau_r^{\alpha-1}}\right)\right\} =\frac{x_0}{\alpha }A^{-1/2\alpha} \int_0^{A} dz \exp\left\{-z\left(1+\frac{\delta_t^\alpha}{t\tau_r^{\alpha-1}}\right)\right\} z^{\frac{1}{2\alpha}-1}.\end{equation}
We can expand the integration boundaries to infinity (the contribution from $A$ to infinity is small as $A\gg 1$). Thus,
\begin{equation} S(t|x_0)\simeq \frac{x_0}{\alpha }A^{-1/2\alpha} \int_0^{\infty} dz \exp\left\{-z\left(1+\frac{\delta_t^\alpha}{t\tau_r^{\alpha-1}}\right)\right\} z^{\frac{1}{2\alpha}-1} =\frac{x_0}{\alpha }A^{-1/2\alpha} \Gamma \left(\frac{1}{2\alpha}\right) \frac{1}{\left(1+\frac{\delta_t^\alpha}{t\tau_r^{\alpha-1}}\right)^{1/2\alpha}}, 
\end{equation}
where in the last equality we identified the Gamma function. Note, in the limit of $t\gg \tau_r$, that $\frac{\delta_t^\alpha}{t\tau_r^{\alpha-1}}\leq \frac{\tau_r^\alpha}{t\tau_r^{\alpha-1}}\leq \frac{\tau_r}{t}\ll 1$. Therefore, as stated in Eq.~\eqref{Survival_Tail_Half-Line}, we obtained that for times much larger than the relaxation time ($t\gg\tau_r$) the survival probability is 
\begin{equation}\label{SI:survival-tail}
S(t|x_0) \simeq \frac{x_0\sqrt{2} \tau_r^{(1-\alpha)/2\alpha}}{\alpha\pi}\Gamma\left(\frac{1}{2\alpha}\right)t^{-1/2\alpha}.
\end{equation}

\begin{figure}[t]
\centering
\includegraphics[width=\textwidth]{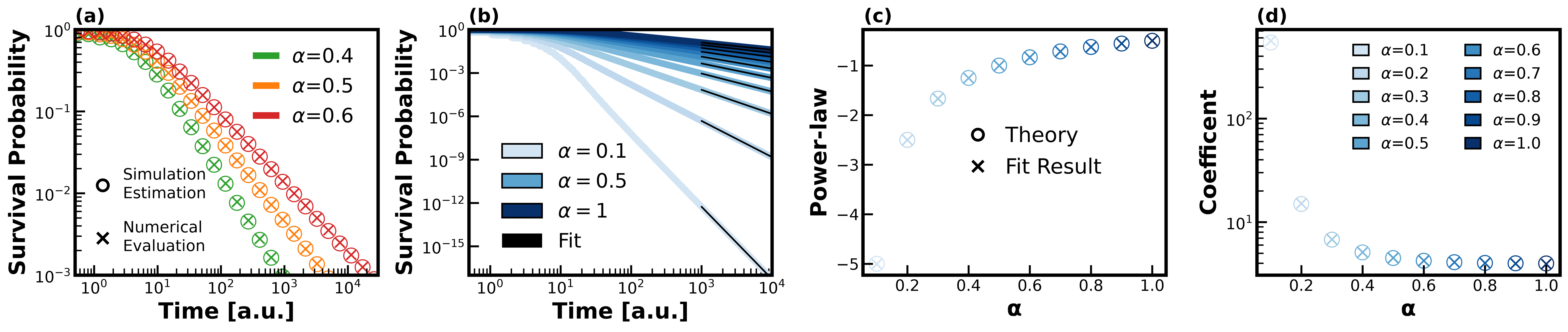}
  \caption{\textbf{Numerical Evaluations vs. Simulation for the Survival Tail.} \textbf{a.} Survival probability as a function of time on a log-log plot. Simulations are shown as empty circles and numerical evaluations as crosses. Comparison is shown for three different values of $\alpha$: $\alpha=0.6$ in red, $\alpha=0.5$ in orange and $\alpha=0.4$ in green. \textbf{b.} Numerical evaluation of the survival probability for numerous values of $\alpha$. Darkest blue corresponds to the highest value, $\alpha=1$ and as the shade lightens the value of $\alpha$ decreases until reaching $\alpha=0.1$. Fits for tail behavior are drawn in black. \textbf{c.} Power-law of the survival tail as a function of $\alpha$. Empty circles indicate the theoretical value ($1/2\alpha$) and crosses are the fit results from (b). \textbf{d.} Comparison of the coefficient of the survival probability tail. Empty circles are the theoretical values $\left(\frac{x_0\sqrt{2} \tau_r^{(1-\alpha)/2\alpha}}{\alpha\pi}\Gamma\left(\frac{1}{2\alpha}\right)\right)$ and crosses are values obtained from the fit in (b).  Across all panels $\tau_r$ was set to 1.  }  \label{Supporting_Survival_Probability}
\end{figure}

In \cref{Supporting_Survival_Probability}(a), we show the survival probability as a function of time for three values of $\alpha$: $\alpha=0.6$ (red), $\alpha=0.5$ (orange) and $\alpha=0.4$ (green). Empty circles and crosses show results from simulations and numerical evaluations, respectively. The two methods match well. In \cref{Supporting_Survival_Probability}(b), we show the numerical evaluation of the survival probability as a function of time for numerous values of $\alpha$, from $\alpha=0.1$ (lightest blue) to $\alpha=1$ (darkest blue). For the tail behavior we perform a fit shown in black.  In \cref{Supporting_Survival_Probability}(c), we show the power-law of the survival probability tail as a function of $\alpha$. Lightest blue corresponds to $\alpha=0.1$ and as the shade darkens $\alpha$ increases until $\alpha=1$.  Empty circles show the theoretical value of the power-law tail, according to \cref{SI:survival-tail}. Crosses show the power-law according to the fit performed in (b). A similar plot is shown in (d), only for the coefficient of the power-law tail behavior. As in (c), the circles indicate the theoretical values and the crosses the fit results. The fit evidently matches the theoretical values for both panels.

\subsection{Mean First Passage Time on the Half-Line (Derivation \cref{MFPT})}\label{Section:MFPT_Half_Line}

To find the MFPT we integrate the survival probability with respect to time, $\text{MFPT}(\alpha,\tau_r, x_0)= \int_0^\infty S(t|x_0)dt$. To solve, we rewrite the integral from zero to infinity as an infinite sum of integrals, 
\begin{equation}\begin{split}
\text{MFPT}(\alpha,\tau_r, x_0)&= \int_0^\infty S(t|x_0)dt=\sum_{n=0}^\infty \int_{n\cdot \tau_r}^{(n+1)\cdot \tau_r} S(t|x_0)dt = \\
&=\sum_{n=0}^\infty \int_{n\cdot \tau_r}^{(n+1)\cdot \tau_r} \int_0^1 \sin(p \pi x_0) \frac{\sin(p\pi )}{1-\cos(p\pi)}  \Tilde{\Lambda}_t(1-\cos(p\pi))   dp =\\
&=\sum_{n=0}^\infty  \int_0^1 dp \sin(p\pi x_0)\frac{\sin(p\pi)}{1-\cos(p\pi)} \int_{n\cdot \tau_r}^{(n+1) \cdot \tau_r} \Tilde{\Lambda}_t (1-\cos(p\pi)) dt.
\end{split}\end{equation}
Next, we plug in the expression of the Laplace transform of the cumulative  rate. Combining \cref{Laplace_Transform_Product} and \cref{ML-Laplace_Transform} gives
\begin{equation}\text{MFPT}(\alpha,\tau_r, x_0)=   \sum_{n=0}^\infty  \int_0^1 dp \sin(p\pi x_0)\frac{\sin(p\pi)}{1-\cos(p\pi)}   \int_{n\cdot \tau_r}^{(n+1) \cdot \tau_r}  \left[\frac{1}{1+[1-\cos\left(p\pi\right)]^\alpha \tau_r^\alpha}\right]^{n_{\tau_r}} \frac{1}{1+[1-\cos\left(p\pi\right)]^\alpha \delta_t^\alpha} dt. \end{equation}
Now, we note that for $n \cdot \tau_r  \leq t \leq  (n+1) \cdot \tau_r$ we have $n_{\tau_r}= n$ and $\delta_t = t-n\tau_r$,
\begin{equation} \text{MFPT}(\alpha,\tau_r, x_0)=   \sum_{n=0}^\infty  \int_0^1 dp \sin(p\pi x_0)\frac{\sin(p\pi)}{1-\cos(p\pi)}  \left[\frac{1}{1+[1-\cos(p\pi)]^\alpha \tau_r^\alpha}\right]^n \int_{n\cdot \tau_r}^{(n+1) \cdot \tau_r}\frac{1}{1+[1-\cos\left(p\pi\right)]^\alpha ( t-n\tau_r)^\alpha} dt .  \end{equation}
Perform an integral variable change $z\equiv t-n\tau_r$,
\begin{equation}\begin{split}
\text{MFPT}(\alpha,\tau_r, x_0)&=  \sum_{n=0}^\infty  \int_0^1 dp \sin(p\pi x_0)\frac{\sin(p\pi)}{1-\cos(p\pi)} \left[\frac{1}{1+[1-\cos(p\pi)]^\alpha \tau_r^\alpha}\right] ^n\int_{0}^{\tau_r}\frac{1}{1+[1-\cos\left(p\pi\right)]^\alpha z^\alpha} dz =\\
&=     \int_0^1 dp \sin(p\pi x_0)\frac{\sin(p\pi)}{1-\cos(p\pi)} \int_{0}^{\tau_r}\frac{1}{1+[1-\cos\left(p\pi\right)]^\alpha z^\alpha} dz \sum_{n=0}^\infty \left[\frac{1}{1+[1-\cos(p\pi)]^\alpha \tau_r^\alpha}\right]^n\\
&=     \int_0^1 dp \sin(p\pi x_0)\frac{\sin(p\pi)}{1-\cos(p\pi)}  \int_{0}^{\tau_r}\frac{1}{1+[1-\cos\left(p\pi\right)]^\alpha z^\alpha} dz  \frac{1}{1- \frac{1}{1+[1-\cos\left(p\pi\right)]^\alpha \tau_r^\alpha}} =\\
&=     \int_0^1 dp \sin(p\pi x_0)\frac{\sin(p\pi)}{1-\cos(p\pi)}  \frac{1+\left[\left(1-\cos p\pi\right)\tau_r\right]^\alpha}{\left[\left(1-\cos p\pi\right)\tau_r\right]^\alpha}  \int_{0}^{\tau_r}\frac{1}{1+[1-\cos\left(p\pi\right)]^\alpha z^\alpha} dz.
\end{split}\end{equation}
Defining $I_1(\alpha, \tau_r,x_0,p) \equiv \sin(p \pi x_0) \frac{\sin(p\pi )}{1-\cos(p\pi)}  \frac{1+\left[\left(1-\cos p\pi\right)\tau_r\right]^\alpha}{\left[\left(1-\cos p\pi\right)\tau_r\right]^\alpha}$ and $I_2(\alpha,\tau_r, p) \equiv \int_{0}^{\tau_r}\frac{dz}{1+[1-\cos\left(p\pi\right)]^\alpha z^\alpha}$ gives Eq.~\eqref{MFPT} in the main text,
\begin{equation} \label{eq:final_MFPT}
\begin{split}
    \text{MFPT}(\alpha,\tau_r, x_0) =& \int_0^1 dp \sin(p\pi x_0)\frac{\sin(p\pi)}{1-\cos(p\pi)}  \frac{1+\left[\left(1-\cos p\pi\right)\tau_r\right]^\alpha}{\left[\left(1-\cos p\pi\right)\tau_r\right]^\alpha}  \int_{0}^{\tau_r}\frac{1}{1+[1-\cos\left(p\pi\right)]^\alpha z^\alpha} dz =\\
    =&\int_0^1 dp I_1(\alpha, \tau_r,x_0,p)\cdot  I_2(\alpha,\tau_r,p).
    \end{split}
\end{equation}

Does this integral diverge or converge? We note first that the integration boundaries of the integrals are finite. Next we make note that in the said the integration boundaries there exists only one pole - for the integral with respect to $p$ at $p=0$. Let us investigate the behavior of the integrand around $p=0$.
\begin{equation} \label{small_p_integrand}
    \begin{split} 
   \sin(p\pi x_0)\frac{\sin(p\pi)}{1-\cos(p\pi)}  \frac{1+\left[\left(1-\cos p\pi\right)\tau_r\right]^\alpha}{\left[\left(1-\cos p\pi\right)\tau_r\right]^\alpha}  \int_{0}^{\tau_r}\frac{1}{1+[1-\cos\left(p\pi\right)]^\alpha z^\alpha} dz &\simeq \frac{2^{\alpha+1}x_0}{\pi^{2\alpha}\tau_r^\alpha} \frac{1}{p^{2\alpha}} \int_0^{\tau_r} dz \\
&=\frac{2^{\alpha+1}x_0}{\pi^{2\alpha}}    \tau_r^{1-\alpha}\frac{1}{p^{2\alpha}}.
\end{split}
\end{equation}
We immediately see the integrand is integrable in the vicinity of $p=0$ if and only if $\alpha<1/2$. Otherwise, it diverges.

\subsubsection{Scaling of Mean First Passage Time with Initial Position (Derivation of \cref{eq:MFPT_x0})} \label{Sec:MFPT_Scaling_x0}

We now investigate the scaling of the MFPT with respect to the initial position $x_0$. Specifically, we are interested in the scaling for $x_0\gg 1$. We  choose some $\varepsilon\in [0,1]$ , sufficiently small such that i) the trigonometric Taylor expansions with respect to $p$ provide good approximations for $p\in[0,\varepsilon]$ and ii) $\frac{(\pi\varepsilon)^{2\alpha}\tau_r^\alpha}{2^\alpha}\ll 1$.  We then split the integral in \cref{eq:final_MFPT} with respect to $p$ into two integrals: the first from 0 to $\varepsilon$, and the second from $\varepsilon$ to 1. We will show the first integral scales like $x_0^{2\alpha}$ while the second as $x_0^{-1}$. Clearly, as $\alpha>0$, the first is the leading order term for $x_0\gg 1$. We will now show these claims explicitly. To start, we focus on the first integral, namely
\begin{equation}  I_{[0,\varepsilon]} \equiv \int_0^{\varepsilon} dp \sin(p\pi x_0)\frac{\sin(p\pi)}{1-\cos(p\pi)}  \frac{1+\left[\left(1-\cos p\pi\right)\tau_r\right]^\alpha}{\left[\left(1-\cos p\pi\right)\tau_r\right]^\alpha}  \int_{0}^{\tau_r}\frac{1}{1+[1-\cos\left(p\pi\right)]^\alpha z^\alpha} dz.\end{equation}
We can approximate the trigonometric terms, that don't depend on $x_0$, using the Taylor approximation. We obtain 
\begin{equation}    I_{[0,\varepsilon]}\simeq  \int_0^{\varepsilon} dp \sin(p\pi x_0)\frac{p\pi}{\frac{(p\pi)^2}{2}}  \frac{1+\frac{(p\pi)^{2\alpha}\tau_r^\alpha}{2^\alpha}}{\frac{(p\pi)^{2\alpha}\tau_r^\alpha}{2^\alpha}}  \int_{0}^{\tau_r}\frac{1}{1+\frac{(p\pi)^{2\alpha}z^\alpha}{2^\alpha}} dz. \end{equation}
Now, we recall that $\varepsilon$ was chosen such that $\frac{(\varepsilon\pi)^{2\alpha}\tau_r^\alpha}{2^\alpha}\ll 1 $. Since $p\leq \varepsilon$ and $z\leq \tau_r$, we have $\frac{(p\pi)^{2\alpha}\tau_r^\alpha}{2^\alpha}\ll 1 $ and $\frac{(p\pi)^{2\alpha}z^\alpha}{2^\alpha}\ll 1 $. Thus we are left with 
\begin{equation}\int_0^{\varepsilon} dp \sin(p\pi x_0)\frac{p\pi}{\frac{(p\pi)^2}{2}}  \frac{1}{\frac{(p\pi)^{2\alpha}\tau_r^\alpha}{2^\alpha}}  \int_{0}^{\tau_r} dz = \int_0^{\varepsilon} dp \sin(p\pi x_0)\frac{2}{p\pi}  \frac{2^\alpha}{(p\pi)^{2\alpha}\tau_r^\alpha}  \tau_r = 2^{\alpha+1}\tau_r^{1-\alpha}\int_0^{\varepsilon} dp   \frac{\sin(p\pi x_0)}{(p\pi)^{2\alpha+1 }}  .\end{equation}
Next we perform an integral variable change $u\equiv p \pi x_0$,  
\begin{equation}2^{\alpha+1}\tau_r^{1-\alpha}\int_0^{\varepsilon} dp   \frac{\sin(p\pi x_0)}{(p\pi)^{2\alpha+1 }} = 2^{\alpha+1}\tau_r^{1-\alpha}\int_0^{\pi x_0 \varepsilon} \frac{du}{\pi x_0}   \frac{\sin(u)}{(u/x_0)^{2\alpha+1 }} = \frac{2^{\alpha+1}\tau_r^{1-\alpha}}{\pi} x_0^{2\alpha}\int_0^{\pi x_0\varepsilon} du   \frac{\sin(u)}{u^{2\alpha+1 }}. \end{equation}
Note that for $x_0\gg \frac{1}{\varepsilon}\gg 1$ we approximate 
\begin{equation}\frac{2^{\alpha+1}\tau_r^{1-\alpha}}{\pi} x_0^{2\alpha}\int_0^{\pi x_0 \varepsilon} du   \frac{\sin(u)}{u^{2\alpha+1 }}\simeq  \frac{2^{\alpha+1}\tau_r^{1-\alpha}}{\pi} x_0^{2\alpha}\int_0^{\infty} du   \frac{\sin(u)}{u^{2\alpha+1 }}.  \end{equation}
The integral left is the Mellin transform of sine and equals $\Gamma(-2\alpha)\sin(-\pi\alpha)$ \cite{Mellin_Transform}. Importantly, since $\alpha<\frac{1}{2}$, then $-1<-2\alpha\leq 0$, or in other words $-2\alpha \not\in \mathbb{Z}$, and so $\Gamma(-2\alpha)$ is finite. We utilize properties of the sine function and Gamma function to yield $\Gamma(-2\alpha)\sin(-\pi\alpha) = \frac{\Gamma(1-2\alpha)}{2\alpha}\sin(\pi\alpha)$. Putting it all together we get 
\begin{equation} I_{[0,\varepsilon]} \simeq \frac{2^{\alpha}\tau_r^{1-\alpha}\Gamma(1-2\alpha)\sin(\pi\alpha)}{\pi\alpha} x_0^{2\alpha}.  \end{equation}

Now we will evaluate the second integral, from $\varepsilon$ to 1, 
\begin{equation}  I_{[\varepsilon,1]} \equiv \int_{\varepsilon}^1 dp \sin(p\pi x_0)\frac{\sin(p\pi)}{1-\cos(p\pi)}  \frac{1+\left[\left(1-\cos p\pi\right)\tau_r\right]^\alpha}{\left[\left(1-\cos p\pi\right)\tau_r\right]^\alpha}  \int_{0}^{\tau_r}\frac{1}{1+[1-\cos\left(p\pi\right)]^\alpha z^\alpha} dz.\end{equation}
Let us denote $I(p,\alpha,\tau_r) = \frac{\sin(p\pi)}{1-\cos(p\pi)}  \frac{1+\left[\left(1-\cos p\pi\right)\tau_r\right]^\alpha}{\left[\left(1-\cos p\pi\right)\tau_r\right]^\alpha}  \int_{0}^{\tau_r}\frac{1}{1+[1-\cos\left(p\pi\right)]^\alpha z^\alpha} dz $, this way 
\begin{equation} I_{[\varepsilon,1]} =  \int_{\varepsilon}^1 dp \sin(p\pi x_0) I(p,\alpha,\tau_r). \end{equation}
We note that for this range, $I(p,\alpha,\tau_r)$ has no singularities, and in particular $\int_{\varepsilon}^1 I(p,\alpha,\tau_r) dp < \infty $. Therefore from the Riemann-Lebesgue lemma we conclude that $I_{[\varepsilon,1]}$ goes to zero as $x_0\shortrightarrow \infty$ \cite{Reimann_Lebesgue}. 

We can also see this explicitly and get a scaling with respect to $x_0$ if we integrate by parts. After integrating by parts, we get
\begin{equation}I(p,\alpha,\tau_r) = \left. \frac{-I(p,\alpha,\tau_r)\cos(p\pi x_0)}{\pi x_0}\right|_{\varepsilon}^1 + \frac{1}{\pi x_0} \int_{\varepsilon}^1 dp \cos(p\pi x_0) \frac{\partial I(p,\alpha,\tau_r)}{\partial p} \sim \mathcal{O}\left(\frac{1}{x_0}\right).\end{equation}

Importantly, as mentioned,  $I(p,\alpha,\tau_r)$ has no singularities at $p=\varepsilon$ or $p=1$ (its only sensitivity is at $p=0$, see discussion in \cref{Section:MFPT_Half_Line}), and thus equals some finite value.  The cosine, naturally, is bounded by 1, thus $\left. \frac{-I(p,\alpha,\tau_r)\cos(p\pi x_0)}{\pi x_0}\right|_{\varepsilon}^1\sim \frac{1}{x_0}$. Similarly, for the second integral, we can either perform a second integration by parts that will yield $\mathcal{O}\left(\frac{1}{x_0^2}\right)$, or note that $\frac{1}{\pi x_0} \int_{\varepsilon}^1 dp \cos(p\pi x_0) \frac{\partial I(p,\alpha,\tau_r)}{\partial p} < \frac{1}{\pi x_0} \int_{\varepsilon}^1 dp  \frac{\partial I(p,\alpha,\tau_r)}{\partial p}  \sim \mathcal{O}\left(\frac{1}{x_0}\right)$.

In sum, as shown in Eq.~\eqref{eq:MFPT_x0},
\begin{equation} \label{eq:MFPT_x_0_sum}
    \text{MFPT}(\alpha<\frac{1}{2},\tau_r, x_0) \simeq \frac{2^{\alpha}\tau_r^{1-\alpha}\Gamma(1-2\alpha)\sin(\pi\alpha)}{\pi\alpha} x_0^{2\alpha} + \mathcal{O}\left(\frac{1}{x_0}\right) \simeq \frac{2^{\alpha}\tau_r^{1-\alpha}\Gamma(1-2\alpha)\sin(\pi\alpha)}{\pi\alpha} x_0^{2\alpha}.
\end{equation}

\begin{figure}[H]
\centering
\includegraphics[width=0.95\textwidth]{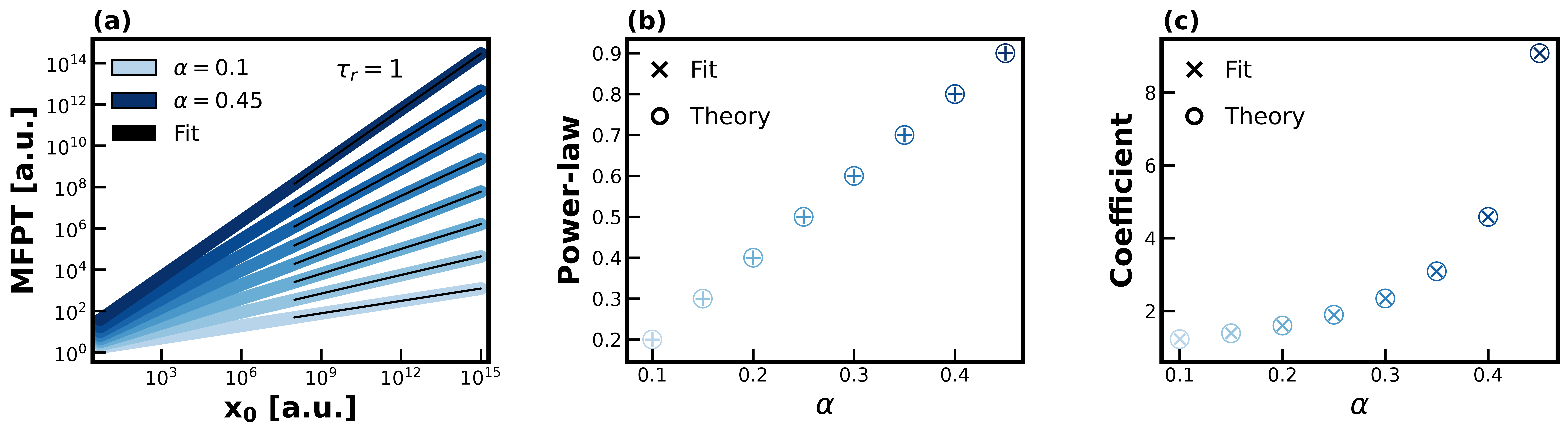}
  \caption{\textbf{Scaling of the MFPT with the Initial Position  $x_0$.}  The MFPT as a function of the initial position $x_0$ for $\tau_r=1.0$, darkest shade corresponds to $\alpha=0.45$ and the shade gets lighter as $\alpha$ decreases. \textbf{a.}  Numerical evaluation of the MFPT \cref{eq:final_MFPT} as a function of $x_0$. Fits are in black.  \textbf{b.} Power-law as a function of $\alpha$. Empty circles indicate the theoretical values, $2\alpha$, and crosses the fit results. \textbf{c.} Coefficient as a function of $\alpha$. Empty circles indicate the theoretical values, $\frac{2^{\alpha}\tau_r^{1-\alpha}\Gamma(1-2\alpha)\sin(\pi\alpha)}{\pi\alpha}$, and crosses the fit results.   }  \label{fig:MFPT_x0_dependency_fit}
\end{figure}

\cref{fig:MFPT_x0_dependency_fit} verifies this relation. In \cref{fig:MFPT_x0_dependency_fit}(a) we show the MFPT as a function of $x_0$, for different values of $\alpha<\frac{1}{2}$  and $\tau_r=1$. The plot is shown in log-log and a clear power-law is exhibited. A fit for each value of $\alpha$ is shown in black. In \cref{fig:MFPT_x0_dependency_fit}(b)-(c) we verify the fit power-law and coefficient with the theoretical values. Both results show good agreement with the predictions in \cref{eq:MFPT_x_0_sum}.

\subsubsection{Mean First Passage Time for $\alpha\shortrightarrow 0$ and  $\alpha\shortrightarrow \frac{1}{2}$}\label{Section:MFPT_Half_Line_Alpha_limits}
We now explore the behavior of the MFPT in \cref{eq:final_MFPT} for $\alpha\shortrightarrow 0$ and  $\alpha\shortrightarrow \frac{1}{2}$. 
Let us start with $\alpha\shortrightarrow 0$. In this limit  $\left[\left(1-\cos p\pi\right)\tau_r\right]^\alpha\simeq 1$, as well as $z^\alpha\simeq 1$. We thus approximate 
\begin{equation}\begin{split}
\text{MFPT}(\alpha\shortrightarrow 0,\tau_r, x_0) & \simeq        \int_0^1 dp \sin(p\pi x_0)\frac{\sin(p\pi)}{1-\cos(p\pi)}  \frac{1+1}{1}  \int_{0}^{\tau_r}\frac{1}{1+1} dz 
=\int_0^1 dp \sin(p\pi x_0)\frac{\sin(p\pi)}{1-\cos(p\pi)}    \int_{0}^{\tau_r} dz=\\
&=\tau_r\int_0^1 dp \sin(p\pi x_0)\frac{\sin(p\pi)}{1-\cos(p\pi)}. 
\end{split}
\end{equation}

We will show that
\begin{equation} \label{Trig_Integral_On_Dirichlet}
    \int_0^1 dp \sin(p\pi x_0)\frac{\sin(p\pi)}{1-\cos(p\pi)} = 1 .
\end{equation}

To prove this, we will show that the  integrand can be written as
\begin{equation} \label{lemma1}
    \sin(p\pi x_0)\frac{\sin(p\pi)}{1-\cos(p\pi)}=\frac{\sin\left(\left(x_0+\frac{1}{2}\right)p\pi\right)}{\sin \left(\frac{p\pi}{2}\right)} - \cos(x_0 p\pi).
\end{equation}

We start by proving the following non-trivial trigonometric identity
\begin{equation} \label{lemma2}
    \sin (nx) \cot \left(\frac{x}{2}\right) = \frac{\sin\left(\left(n+\frac{1}{2}\right)x\right)}{\sin \left(\frac{x}{2}\right)} - \cos(nx),
\end{equation}
for $n\in\mathbb{N}$, and $x\in \mathbb{R}$. 
Recall that $\sin(a+b)=\sin(a)\cos(b)+\cos(a)\sin(b)$ \cite{Trig_Treat}, thus
\begin{equation}\sin\left(\left(n+\frac{1}{2}\right)x\right) = \sin(nx)\cos \left(\frac{x}{2}\right) + \cos(nx)\sin\left(\frac{x}{2}\right).\end{equation}
Divide by $\sin\left(\frac{x}{2}\right)$ and we get the required,
\begin{equation}\frac{\sin\left(\left(n+\frac{1}{2}\right)x\right)}{\sin\left(\frac{x}{2}\right)} = \frac{\sin(nx)\cos \left(\frac{x}{2}\right)}{\sin \left(\frac{x}{2}\right)} + \frac{\cos(nx)\sin\left(\frac{x}{2}\right)}{\sin\left(\frac{x}{2}\right)} = \sin(nx) \cot \left(\frac{x}{2}\right) + \cos(nx). \end{equation}
Now we are ready to prove \cref{lemma1}. First use the trigonometric identity $\cot \frac{x}{2} = \frac{\sin x}{1-\cos x}$ \cite{Trig_Treat} for $x= p \pi$,
\begin{equation}\int_0^1 dp \sin(p\pi x_0)\frac{\sin(p\pi)}{1-\cos(p\pi)} = \int_0^1 dp \sin (p\pi x_0)  \cot \left(\frac{p\pi}{2}\right). \end{equation}
Now we use \cref{lemma2},
\begin{equation} \int_0^1 dp \sin (p\pi x_0)  \cot \left(\frac{p\pi}{2}\right)= \int_0^1 dp \frac{\sin\left(\left(x_0+\frac{1}{2}\right)p\pi\right)}{\sin \left(\frac{p\pi}{2}\right)} - \cos(x_0 p\pi) . \end{equation}
The cosine term drops (one can see this either by integrating explicitly, or by changing integration variable to $u=p\pi$ and noticing the integral has symmetric boundaries and the integrand is an odd function), and we are left with
\begin{equation}  \int_0^1 dp \frac{\sin\left(\left(x_0+\frac{1}{2}\right)p\pi\right)}{\sin \frac{p\pi}{2}} . \end{equation}
We are now ready to use the Dirichlet kernel \cite{Dirichlet_Proof,DirichletKerneProof2}, $\frac{\sin\left(\left(n+1/2\right)x\right)}{\sin \left(x/2\right)}  =  1+2\overset{n}{\underset{k=1}{\sum}}\cos(k x)$, for $x=p\pi$ and $n=x_0$, to rewrite the sine ratio in the integral, 
\begin{equation}  \int_0^1 dp \frac{\sin\left(\left(x_0+\frac{1}{2}\right)p\pi\right)}{\sin \left(\frac{p\pi}{2}\right)}  = \int_0^1 \: dp 1+2\overset{x_0}{\underset{k=1}{\sum}}\cos(k p\pi)  .\end{equation}
We now integrate, all cosine terms drop and we are left with 1,
\begin{equation}\int_0^1 dp \: 1+2\overset{x_0}{\underset{k=1}{\sum}}\cos(p\pi) = 1. \end{equation}
Therefore, we get that for all values of $\tau_r$, when $\alpha$ approaches zero the MFPT is
\begin{equation}
    \text{MFPT} \label{MFPT_alphazrero} (\alpha\shortrightarrow 0,\tau_r, x_0) \simeq \tau_r.
\end{equation}

There is an intuitive reasoning for the result. In this limit, the Mittag-Leffler distribution becomes increasingly concentrated at two extremes: rate zero and rate infinity (we will show this explicitly momentarily). The probability of sampling either converges to half. Thus, sampling a rate is analogous to flipping a coin: heads for rate zero; tails for rate infinity.  When we flip tails (rate infinity), the walker reaches the target in zero time. This occurs on average after two flips. As we first flip at $t=0$, and then again every $\tau_r$,  the second flip is at $t=\tau_r$. Therefore, the MFPT is  $\tau_r$. 

We will now prove that in the limit $\alpha\shortrightarrow 0$ the Mittag-Leffler distribution becomes concentrated at two extremes: zero and infinity. Recall that the survival function of the Mittag-Leffler distribution is defined from the Mittag-Leffler function \cite{Mittag_Leffler_Pillai,Mittag_Leffler_Asymptotics,gorenflo2010mittagleffler}, namely 
\begin{equation}P(\lambda>x) = E_\alpha(-x^\alpha),\end{equation}
where the Mittag-Leffler function, for $x>0$  is \cite{Mittag_Leffler_Function_Integral_Form} 
\begin{equation} \label{ML_Function_Integral}
    E_\alpha(-x^\alpha) =  \int_0^\infty \frac{1}{\pi\alpha} e^{-r^{1/\alpha}}\frac{x^\alpha\sin(\pi\alpha)}{r^2+2rx^\alpha\cos(\pi\alpha)+x^{2\alpha}}dr.
\end{equation}
We thus have
\begin{equation}P(\lambda>x)  = \int_0^\infty \frac{1}{\pi\alpha} e^{-r^{1/\alpha}}\frac{x^\alpha \sin(\pi\alpha)}{r^2+2rx^\alpha\cos(\pi\alpha)+x^{2\alpha}}dr =\frac{\sin(\pi\alpha)}{\pi\alpha}x^\alpha\int_0^\infty\frac{e^{-r^{1/\alpha}}}{r^2+2rx^\alpha\cos(\pi\alpha)+x^{2\alpha}}dr .\end{equation}
Now, we take the limit $\alpha\shortrightarrow 0$,
\begin{equation}\lim_{\alpha\shortrightarrow 0}P(\lambda>x)  = \lim_{\alpha\shortrightarrow 0}\frac{\sin(\pi\alpha)}{\pi\alpha}x^\alpha\int_0^\infty\frac{e^{-r^{1/\alpha}}}{r^2+2rx^\alpha\cos(\pi\alpha)+x^{2\alpha}}dr =  \int_0^1 \frac{1}{(r+1)^2}dr.\end{equation}
 In the last equality we used $\underset{\alpha\shortrightarrow 0}{\lim}\frac{\sin(\pi\alpha)}{\pi\alpha}\rightarrow 1$ \cite{sinc_limit}, and $x^\alpha\shortrightarrow 1$, in addition note that when $\alpha\shortrightarrow 0$ the numerator in the integrand, $e^{-r^{1/\alpha}}$ equals one when $|r|<1$, and is zero otherwise. After integration the integral equals $\frac{1}{1+1} = \frac{1}{2}$, and thus 
 \begin{equation}\lim_{\alpha\shortrightarrow 0}P(\lambda>x)  =  \frac{1}{2}.\end{equation}

\noindent We have shown that in the limit $\alpha\shortrightarrow 0$, the Mittag-Leffler survival function equals $\frac{1}{2}$  for any positive rate, which implies an atom at infinity. Since $P(\lambda\geq 0)=1$ from normalization, the probability of sampling rate zero is one-half, and similarly for rate infinity. 

Next, we turn our attention to the MFPT for $\alpha\shortrightarrow \frac{1}{2}$. Recall, for $\alpha\geq \frac{1}{2}$ the MFPT diverges. To assess the manner of divergence we define $\Delta\equiv \frac{1}{2}-\alpha$ and we ask how does the MFPT depends on $\Delta$ as $\Delta\shortrightarrow 0$. The divergence of the MFPT is rooted in the behavior of the integrand near $p=0$ (see \cref{Section:MFPT_Half_Line}). We split the MFPT in \cref{eq:final_MFPT} into two integrals with respect to $p$, the first from 0 to $\varepsilon$, $0<\varepsilon<1$, such that $\sin \varepsilon \pi \simeq \varepsilon \pi$, and the second from $\varepsilon$ to 1. We use our previous \cref{small_p_integrand} to write, 
\begin{equation} 
\begin{split}
\text{MFPT}
(\alpha = \frac{1}{2}-\Delta,\tau_r, x_0) &\simeq \frac{2^{\alpha+1}x_0}{\pi^{2\alpha}}    \tau_r^{1-\alpha}\int_0^\varepsilon\frac{dp}{p^{2\alpha}}+\\
&\phantom{\simeq}+\int_\varepsilon^1 dp \sin(p\pi x_0)\frac{\sin(p\pi)}{1-\cos(p\pi)}  \frac{1+\left[\left(1-\cos p\pi\right)\tau_r\right]^\alpha}{\left[\left(1-\cos p\pi\right)\tau_r\right]^\alpha}  \int_{0}^{\tau_r}\frac{1}{1+[1-\cos\left(p\pi\right)]^\alpha z^\alpha} dz.
\end{split} 
\end{equation}
Next, as $\alpha\shortrightarrow \frac{1}{2}$, or $\Delta\shortrightarrow 0$, the divergence comes from the first integral, and so
the second integral is negligible compared to it. In this limit we have
\begin{equation} \text{MFPT}(\alpha = \frac{1}{2}-\Delta,\tau_r, x_0) \simeq \frac{2^{\alpha+1}x_0}{\pi^{2\alpha}}    \tau_r^{1-\alpha}\int_0^\varepsilon\frac{dp}{p^{2\alpha}} =\frac{2^{\alpha+1}x_0}{\pi^{2\alpha}} \tau_r^{1-\alpha} \frac{\varepsilon^{1-2\alpha}}{1-2\alpha} \sim \frac{1}{1-2\alpha} .  \end{equation}
 In terms of $\Delta$, we have 
 \begin{equation} \label{MFPT_alphahalf}
      \text{MFPT}(\Delta = \frac{1}{2}-\alpha,\tau_r, x_0) \sim \frac{1}{\Delta}.
 \end{equation}
Most importantly, when $\Delta$ approaches zero, the divergence is of the order of magnitude of $\frac{1}{\Delta}$.

\begin{figure}[t!]
\centering
\includegraphics[width=0.9\textwidth]{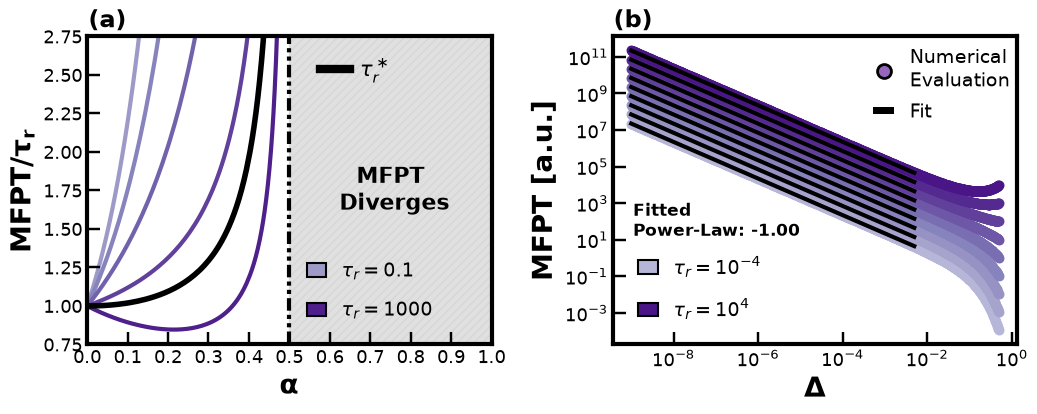}
  \caption{\textbf{MFPT  Behavior for $\alpha\shortrightarrow 0$ and $\alpha\shortrightarrow \frac{1}{2}$.} \textbf{a.} The MFPT, normalized by $\tau_r$, as a function of $\alpha$. Values of $\tau_r$ vary from $\tau_r=0.1$ (lightest purple) to $\tau_r=1000$ (darkest purple). Critical $\tau_r$ is in black. \textbf{b.} The MFPT as a function of $\Delta\equiv \frac{1}{2}-\alpha$.  Values of $\tau_r$ vary from $\tau_r=10^{-4}$ (lightest purple) to $\tau_r=10^4$ (darkest purple). Fit for $\Delta\ll 1$ is in black, slope for all fits is -1.   }  \label{Supporting_Survival_MFPT_alphas}
\end{figure}

The results of this section are illustrated in \cref{Supporting_Survival_MFPT_alphas}. In \cref{Supporting_Survival_MFPT_alphas}(a) the MFPT, normalized by the relaxation time $\tau_r$, is plotted as a function of $\alpha$, for different values of $\tau_r$. As $\tau_r$ increases, so does the shade of purple. The value at the critical relaxation time is in black. For $\tau_r < \tau_r^\ast$ the MFPT monotonically increases, and for $\tau_r >\tau_r^\ast$ the MFPT dips after $\alpha=0$ and there is a minimum, until it increases again and diverges at $\alpha = \frac{1}{2}$  (see main text, proof in the next section  \cref{Section:MFPT_Half_Line_Minimum}). Additionally, we observe that for all values of $\tau_r$, at $\alpha\shortrightarrow 0$, the MFPT equals $\tau_r$ \cref{MFPT_alphazrero}. To inspect the behavior of the MFPT on the other regime, $\alpha\shortrightarrow \frac{1}{2}$, we turn our attention to \cref{Supporting_Survival_MFPT_alphas}(b). There, the MFPT is shown as a function of $\Delta\equiv \frac{1}{2}-\alpha$. Several values of $\tau_r$ are plotted, from $\tau_r=10^{-4}$ (lightest purple) to $\tau_r = 10^{4} $ (darkest purple). A fit for all values of $\tau_r$ was performed for $\Delta \ll 1$, and all have the same power-law of -1, as predicted in \cref{MFPT_alphahalf}.

\subsubsection{The Critical Relaxation Time $\tau_r^*$} \label{Section:MFPT_Half_Line_Minimum}
Here, we find the critical relaxation time $\tau_r^*$ beyond which the MFPT has a minimum as a function of $\alpha$. We will show that at $\alpha=0$ the MFPT changes from increasing to decreasing, at some critical $\tau_r$. From this we  conclude that there is at least one minimum, since the MFPT diverges as $\alpha\to1/2$. 

To discover if the MFPT increases or decreases at $\alpha=0$ we derive the MFPT \cref{eq:final_MFPT} with respect to $\alpha$. The derivative is 
\begin{equation}\begin{split}\frac{\partial\text{MFPT}(\alpha,\tau_r, x_0)}{\partial \alpha}&=   \int_0^1 dp \sin(p\pi x_0) \frac{\sin p\pi}{1-\cos p\pi} \int_0^{\tau_r}\frac{dz}{1+(1-\cos p\pi)^\alpha z^\alpha} \left\{-\frac{\ln\left(1-\cos p\pi\right) \left[1+(1-\cos p\pi)^{\alpha} \tau_r^{\alpha}\right]}{(1-\cos p \pi)^{\alpha} \tau_r^{\alpha}}\right. - \\
&-\frac{\ln\tau_r \left[1+(1-\cos p\pi)^{\alpha} \tau_r^{\alpha}\right]}{(1-\cos p \pi)^{\alpha} \tau_r^{\alpha}}+\ln(1-\cos \pi p)  +  \ln\tau_r -   \frac{z^\alpha (1-\cos p\pi)^{\alpha} \ln(1-\cos p\pi)\left[1+(1-\cos p\pi)^{\alpha} \tau_r^{\alpha} \right] }{(1-\cos p\pi)^{\alpha} \tau_r^{\alpha} \left[1+(1-\cos p\pi)^{\alpha} z^{\alpha}\right]} - \\
&-\left.\frac{z^{\alpha}\ln z(1-\cos p\pi)^\alpha\left[1+(1-\cos p\pi)^{\alpha} \tau_r^{\alpha} \right]}{(1-\cos p\pi)^{\alpha} \tau_r^{\alpha} \left[1+(1-\cos p\pi)^{\alpha} z^{\alpha}\right]}\right\}.\end{split}\end{equation}

Now we take the limit $\alpha\shortrightarrow 0$.  In this limit  $\left(1-\cos p\pi\right)^\alpha\simeq 1$, $\tau_r^\alpha\simeq 1$, and  $z^\alpha\simeq 1$, we obtain 

\begin{equation}\begin{split}\lim_{\alpha\shortrightarrow 0}\frac{\partial\text{MFPT}(\alpha,\tau_r, x_0)}{\partial \alpha} &=   \int_0^1 dp \sin(p\pi x_0) \frac{\sin p\pi}{1-\cos p\pi} \int_0^{\tau_r}\frac{dz}{1+1} \left\{-\frac{\ln\left(1-\cos p\pi\right) \left[1+1\right]}{1}\right. - \\
&\phantom{=}\left.-\frac{\ln\tau_r \left[1+1\right]}{1}+\ln(1-\cos \pi p)  +  \ln\tau_r -   \frac{1\cdot \ln(1-\cos p\pi)\left[1+1 \right] }{1\cdot \left[1+1\right]} -\frac{1\cdot \ln z\cdot 1 \left[1+1 \right]}{1\cdot \left[1+1\right]}\right\}=\\
&= \int_0^1 dp \sin(p\pi x_0) \frac{\sin p\pi}{1-\cos p\pi} \int_0^{\tau_r}\frac{dz}{1+1} \cdot \left\{-2\ln\left(1-\cos p\pi\right)-2\ln\tau_r + \ln(1-\cos p\pi) + \right. \\ 
&\phantom{=}+\left.\ln\tau_r -\ln(1-\cos p\pi) -\ln z\right\}= \\
&=\int_0^1 dp \sin(p\pi x_0) \frac{\sin p\pi}{1-\cos p\pi} \int_0^{\tau_r}dz \left\{-\ln\left(1-\cos p\pi\right)-\frac{1}{2}\ln\tau_r  -\frac{1}{2}\ln z\right\}.\end{split}\end{equation}
We now integrate with respect to $z$
\begin{equation}
\begin{split}\lim_{\alpha\shortrightarrow 0}\frac{\partial\text{MFPT}(\alpha,\tau_r, x_0)}{\partial \alpha}&=\int_0^1 dp \sin(p\pi x_0) \frac{\sin p\pi}{1-\cos p\pi}\left\{-\tau_r\ln(1-\cos p\pi) -\frac{1}{2}\tau_r\ln\tau_r - \frac{1}{2}\tau_r\ln\tau_r+\frac{1}{2}\tau_r\right\}=\\
&=-\tau_r\int_0^1 dp \sin(p\pi x_0) \frac{\sin p\pi}{1-\cos p\pi}\left\{\ln(1-\cos p\pi) +\ln\tau_r -\frac{1}{2}\right\}. \end{split}\end{equation}
As shown in \cref{Section:MFPT_Half_Line_Alpha_limits}, \cref{Trig_Integral_On_Dirichlet}, for the last two terms, the integral with respect to $p$ equals 1 .Thus we obtain
\begin{equation} \label{derivative_integral}
\lim_{\alpha\shortrightarrow 0}\frac{\partial\text{MFPT}(\alpha,\tau_r, x_0)}{\partial \alpha}= - \tau_r\cdot\left(\int_0^1 dp \sin(p\pi x_0) \frac{\sin p\pi}{1-\cos p\pi}\ln(1-\cos p\pi)\right) -\tau_r\left(\ln\tau_r -\frac{1}{2}\right).\end{equation}
We will now focus our effort on the remaining integral, with respect to $p$. We use \cref{lemma1} (\cref{Section:MFPT_Half_Line_Alpha_limits}) to get 
\begin{equation}\sin(p\pi x_0) \frac{\sin p\pi}{1-\cos p\pi} = \frac{\sin\left(\left(x_0+\frac{1}{2}\right)p\pi\right)}{\sin \left(\frac{p\pi}{2}\right)} - \cos(x_0 p\pi).  \end{equation}
Next, as in \cref{Section:MFPT_Half_Line_Alpha_limits}, we use the Dirichlet kernel \cite{Dirichlet_Proof,DirichletKerneProof2} to obtain 
\begin{equation} \label{lemma_dirichlet}
\frac{\sin\left(\left(x_0+\frac{1}{2}\right)p\pi\right)}{\sin \left(\frac{p\pi}{2}\right)} - \cos(x_0 p\pi) = 1 + 2 \overset{x_0}{\underset{k=1}{\sum}}\cos(kp\pi)  - \cos(x_0 p\pi).\end{equation}

\noindent And so, with the help of \cref{lemma_dirichlet}, the integral in \cref{derivative_integral} can be written as
\begin{equation} \label{integral_p_dirichlet}\int_0^1 dp \ln(1-\cos p\pi) + 2\overset{x_0}{\underset{k=1}{\sum}}\int_0^1 dp \cos(kp\pi)\ln(1-\cos p\pi) -\int_0^1dp\cos(x_0p\pi)\ln(1-\cos p\pi) .\end{equation}

The first integral in \cref{integral_p_dirichlet} simply gives $\ln(2)-\ln(4)=-\ln(2)$. For the next two integrals, we make the following claim
\begin{equation} \label{claim}
    \int_0^1 \cos(k p\pi) \ln(1-\cos p\pi) = -\frac{1}{k} 
\end{equation}
for $k\in \mathbb{N}$. We will now prove this claim explicitly. We start with integration by parts 
\begin{equation} 
\begin{split}
\int_0^1 dp \cos(kp\pi ) \ln(1-\cos p\pi) &= \frac{1}{k\pi}\sin(kp\pi)\ln(1-\cos p\pi)\Big|_0^1  - \frac{1}{k}\int_0^\pi  dp\frac{\sin(kp\pi)\sin p\pi}{1-\cos (p\pi)} =\\
&=- \frac{1}{k}\int_0^1  \sin(kp\pi)dp\frac{\sin (p\pi)}{1-\cos (p\pi)} . \end{split}\end{equation}
\noindent Now, we use \cref{Trig_Integral_On_Dirichlet}, for $k=x_0$,  to obtain
\begin{equation} 
\int_0^1 dp \cos(kp\pi ) \ln(1-\cos p\pi) =  - \frac{1}{k}\int_0^1  \sin(kp\pi)dp\frac{\sin (p\pi)}{1-\cos (p\pi)} = -\frac{1}{k} \cdot 1 = -\frac{1}{k},\end{equation}
as required.
Thus \cref{integral_p_dirichlet} becomes,
\begin{equation} \label{final_p_integral_dirichlet}
\begin{split}
\int_0^1 dp \ln(1-\cos p\pi) + \overset{x_0}{\underset{k=1}{\sum}}\int_0^1 dp \cos(kp\pi)\ln(1-\cos p\pi) -\int_0^1dp\cos(x_0p\pi)\ln(1-\cos p\pi)  &=-\ln(2) - 2\overset{x_0}{\underset{k=1}{\sum}}\frac{1}{k} +\frac{1}{x_0}  = \\
   &=-\ln(2) - 2 H_{x_0} +\frac{1}{x_0}, 
\end{split}\end{equation}
where in the last equality we identified  the sum $\overset{x_0}{\underset{k=1}{\sum}}\frac{1}{k}$ as the $x_0$-th Harmonic number \cite{HarmonicNumber1, HarmonicNumber2, HarmonicNumber3} $H_{x_0}$. In sum, 
\begin{equation} \label{final_p_integral_dirichlet_sum}
    \int_0^1 dp \sin(p\pi x_0) \frac{\sin p\pi}{1-\cos p\pi}\ln(1-\cos p\pi) =  -\ln(2) - 2 H_{x_0} +\frac{1}{x_0}
\end{equation}
Plugging back  the integral \cref{final_p_integral_dirichlet_sum} into \cref{derivative_integral} gives,
\begin{equation}\label{MFPT_Derivative}
\begin{split}\lim_{\alpha\shortrightarrow 0}\frac{\partial\text{MFPT}(\alpha,\tau_r, x_0)}{\partial \alpha}&=-\tau_r\left(-\ln(2) - 2H_{x_0} + \frac{1}{x_0}\right) - \tau_r \left(\ln\tau_r -\frac{1}{2}\right)=\\
&=-\tau_r \left\{-\ln(2) - 2H_{x_0} + \frac{1}{x_0}  + \ln\tau_r -\frac{1}{2} \right\}. \end{split}\end{equation}
We emphasize that no approximations were used, \cref{MFPT_Derivative} is exact.  

We now ask when the derivative of the MFPT with respect to $\alpha$, at $\alpha=0$, changes sign. To find out, we equate the derivative to zero and get the following condition 
\begin{equation}-\ln(2) -2 H_{x_0} + \frac{1}{x_0}  + \ln\tau_r -\frac{1}{2} \overset{!}{=}0.\end{equation}
\noindent Solving for $\tau_r$, we find that the MFPT changes sign at 
\begin{equation} \label{min_condition}
    \tau_r^\ast = \exp\left\{ \ln(2) +2H_{x_0}-\frac{1}{x_0}+\frac{1}{2} \right\}.
\end{equation}

\noindent In sum,
\begin{equation}
\lim_{\alpha\to 0}\frac{\partial\,\text{MFPT}(\alpha,\tau_r, x_0)}{\partial \alpha}
\begin{cases}
> 0 & \tau_r < \tau_r^*, \\
=0 & \tau_r=\tau_r^\ast,
\\
< 0 & \tau_r > \tau_r^*.
\end{cases}
\end{equation}
\noindent Thus, we know that for $\tau_r>\tau_r^\ast$, the MFPT has at least one minimum, which we find empirically to be the only minimum. 

\smallskip

We can also obtain the dependence on $x_0$ from \cref{min_condition}. The Harmonic number can be approximated by $H_{x_0}\simeq \log x_0 + \gamma+\frac{1}{2x_0} $, where $\gamma$ is the Euler-Mascheroni number \cite{HarmonicNumber3}. Plugging this into \cref{min_condition} we get that the critical relaxation time is 
\begin{equation}\begin{split}\tau_r^\ast &\simeq \exp\left\{\ln(2) + 2\log x_0 +\frac{1}{x_0}+2\gamma - \frac{1}{x_0}+\frac{1}{2}\right\} = \exp\left\{2\log x_0+ \ln(2)+2\gamma +\frac{1}{2}\right\}= \\
&=x_0^2 \cdot \exp\left( \ln(2)+2\gamma +\frac{1}{2}\right) \equiv A_\gamma x_0^2 \propto x_0^2, \end{split}\end{equation}
with $A_\gamma \equiv \ln(4) - \ln(2)+2\gamma +\frac{1}{2} $. Note that this approximation is more accurate as $x_0$ increases (for small $x_0$, other terms in the Harmonic number approximation need to be taken into consideration).

\subsubsection{Mean First Passage Time for Small and Large $\tau_r$ (Derivation of \cref{eq:MFPT_Asym})}\label{Section:MFPT_Half_Line_Tau_r_limits}

We now examine the behavior of the MFPT with respect to $\tau_r$, the relaxation time, in the limits of small and large $\tau_r$. We point out that a large relaxation time corresponds to rare changes in the jump rate, whereas a small relaxation time corresponds to frequent changes in the jump rate. We naturally focus only on $\alpha<\frac{1}{2}$, where the MFPT does not diverge.

We inspect \cref{eq:final_MFPT} and pinpoint its two dependencies on $\tau_r$: the first is in the term $\left[\left(1-\cos p\pi\right)\tau_r\right]^\alpha$, and the second is in the upper integration boundary of the integral with respect to $z$. Let us start with small relaxation times, and specifically demand $\tau^\alpha_r\ll 1$ such that $\left[\left(1-\cos p\pi\right)\tau_r\right]^\alpha \ll 1$ for every $p\in[0,1]$. Moreover, since $z\leq \tau_r$, we also have $\left[\left(1-\cos p\pi\right)z\right]^\alpha \ll 1$. Thus, in the small relaxation time limit \cref{eq:final_MFPT} becomes
\begin{equation}\begin{split}
\text{MFPT}(\alpha<\tfrac{1}{2},\tau_r \ll 1, x_0)&=\int_0^1 dp \sin(p \pi x_0) \frac{\sin(p\pi )}{1-\cos(p\pi)}  \frac{1+\left[\left(1-\cos p\pi\right)\tau_r\right]^\alpha}{\left[\left(1-\cos p\pi\right)\tau_r\right]^\alpha} \int_{0}^{\tau_r}\frac{dz}{1+[1-\cos\left(p\pi\right)]^\alpha z^\alpha} \simeq \\
&\simeq \int_0^1 dp \sin(p \pi x_0) \frac{\sin(p\pi )}{1-\cos(p\pi)}  \frac{1}{\left[\left(1-\cos p\pi\right)\tau_r\right]^\alpha} \int_{0}^{\tau_r}\frac{dz}{1}= \\
&=\int_0^1 dp \sin(p \pi x_0) \frac{\sin(p\pi )}{1-\cos(p\pi)}  \frac{1}{\left[\left(1-\cos p\pi\right)\tau_r\right]^\alpha} \tau_r=\\
&=\tau_r^{1-\alpha}  \int_0^1 dp \sin(p \pi x_0) \frac{\sin(p\pi )}{\left[1-\cos(p\pi)\right]^{1+\alpha}}.
\end{split}\end{equation}
 
\noindent In sum, when the relaxation time is small we obtain
\begin{equation} \label{MFPT_Small_Taur_PowerLaw}
    \text{MFPT}(\alpha<\tfrac{1}{2},\tau_r \ll 1, x_0) \simeq \tau_r^{1-\alpha}  \int_0^1 dp \sin(p \pi x_0) \frac{\sin(p\pi )}{\left[1-\cos(p\pi)\right]^{1+\alpha}}  \equiv C_0\,\tau_r^{1-\alpha},
\end{equation}
where we have defined
\begin{equation} \label{C0_definition}
    C_0 \equiv \int_0^1 dp \,\sin(p\pi x_0)  \frac{\sin(p\pi)}{(1-\cos p\pi)^{1+\alpha}} .
\end{equation}

We note in passing that $C_0$ is finite precisely when $\alpha<\frac{1}{2}$. Indeed, near $p=0$ we may expand $\sin(p\pi x_0)\simeq p\pi x_0$, $\sin(p\pi)\simeq p\pi$ and $1-\cos(p\pi)\simeq \frac{(p\pi)^2}{2}$, so that the integrand of \cref{C0_definition} behaves as
\begin{equation} \label{C0_small_p}
   \sin(p\pi x_0)  \frac{\sin(p\pi)}{(1-\cos p\pi)^{1+\alpha}} \simeq \frac{2^{1+\alpha}x_0}{\pi^{2\alpha}}\,\frac{1}{p^{2\alpha}},
   \qquad p\ll 1,
\end{equation}
which is integrable if and only if $2\alpha<1$. This is the same small-$p$ divergence that governs the first-passage transition in \cref{small_p_integrand}.
  
We now turn to the limit of large relaxation times, where more care is required. Naively, we would like to approximate $\left[\left(1-\cos p\pi\right)\tau_r\right]^\alpha \gg 1$ and proceed as above. Yet at $p=0$ we have $\left(1-\cos p\pi\right)=0$, so $\left[\left(1-\cos p\pi\right)\tau_r\right]^\alpha$ is \textit{not} large there, no matter how large $\tau_r$ is. Since $1-\cos(p\pi)$ increases monotonically on $p\in[0,1]$, the region where the approximation fails is an interval $[0,\varepsilon]$ adjacent to the origin. We therefore split the integral in \cref{eq:final_MFPT} into two parts, 
\begin{equation} \label{MFPT_split}
    \text{MFPT}(\alpha<\tfrac{1}{2},\tau_r, x_0) = \mathcal{M}_{[0,\varepsilon]}+\mathcal{M}_{[\varepsilon,1]},
    \qquad
    \mathcal{M}_{[a,b]} \equiv \int_a^b dp\, I_1(\alpha,\tau_r,x_0,p)\cdot I_2(\alpha,\tau_r,p),
\end{equation}
with $I_1$ and $I_2$ as defined below \cref{eq:final_MFPT}, and we let the cutoff depend on the relaxation time, $\varepsilon = \varepsilon(\tau_r)$. Two requirements dictate how fast $\varepsilon(\tau_r)$ may shrink:
\begin{enumerate}

\item[(i)] $\varepsilon\ll 1$, so that the discarded interval $[0,\varepsilon]$ contributes negligibly. 

\item[(ii)] $\varepsilon^{2\alpha}\tau_r^\alpha\gg 1$, so that $A\equiv\left[\left(1-\cos p\pi\right)\tau_r\right]^\alpha\gg 1$ holds \textit{uniformly} on $[\varepsilon,1]$. Indeed, for $p\geq \varepsilon$ monotonicity gives $\left(1-\cos p\pi\right)\tau_r \geq \left(1-\cos \varepsilon\pi\right)\tau_r \simeq \frac{(\pi\varepsilon)^2}{2}\tau_r$, so that $A \geq \left[\frac{(\pi\varepsilon)^2} {2}\tau_r\right]^\alpha\gg1$.

\end{enumerate}

\noindent Requirements (i) and (ii) leave a wide window for the cutoff. Both are met, for instance, by
\begin{equation} \label{epsilon_window}
    \varepsilon(\tau_r) = \frac{1}{\tau_r^{1/\beta}}
    \qquad \text{with} \qquad \beta>2 ,
\end{equation}
for which $\varepsilon\ll1$ and $\varepsilon^{2\alpha}\tau_r^{\alpha}=\tau_r^{\alpha\left(1-2/\beta\right)}\gg 1$ at large $\tau_r$. The window of admissible cutoffs is thus the same for every $0<\alpha<\frac{1}{2}$, although the $\tau_r$ at which requirement (ii) is met is governed by the exponent $\alpha\left(1-2/\beta\right)$. As we will see, the final result does not depend on which admissible $\varepsilon$ is chosen. We now show that $\mathcal{M}_{[0,\varepsilon]}$ vanishes relative to $\tau_r^{1-\alpha}$, while $\mathcal{M}_{[\varepsilon,1]}\sim\tau_r^{1-\alpha}$, yielding the required result.

\textit{The interval $[0,\varepsilon]$.} We do not need to evaluate this contribution, only to bound it. To that end we bound $I_2$ in two complementary ways. Using $\frac{1}{1+w}\leq 1$ and $\frac{1}{1+w}\leq \frac{1}{w}$ for any positive $w$, in this case $w\equiv \left[\left(1-\cos p\pi\right)z\right]^\alpha\geq 0$, we obtain, respectively,
\begin{equation} \label{I2_bounds}
    I_2 = \int_0^{\tau_r}\frac{dz}{1+\left[\left(1-\cos p\pi\right)z\right]^\alpha}
    \;\leq\; \int_0^{\tau_r} dz = \tau_r,
    \qquad
    I_2 \;\leq\; \int_0^{\tau_r} \frac{dz}{\left[\left(1-\cos p\pi\right)z\right]^\alpha}
    = \frac{1}{1-\alpha}\,\frac{\tau_r}{\left[\left(1-\cos p\pi\right)\tau_r\right]^{\alpha}} .
\end{equation}
With $A\equiv \left[\left(1-\cos p\pi\right)\tau_r\right]^\alpha$ as above, the $\tau_r$-dependent factor of $I_1$ is $\frac{1+A}{A}$. If $A\geq 1$ then $\frac{1+A}{A}\leq 2$ and we use the second bound in \cref{I2_bounds}; if $A<1$ then $\frac{1+A}{A}\leq \frac{2}{A}$ and we use the first. Either way,
\begin{equation} \label{integrand_bound}
    \frac{1+A}{A}\,I_2 \;\leq\; \frac{2}{1-\alpha}\,\frac{\tau_r}{A} = \frac{2}{1-\alpha}\,\frac{\tau_r^{1-\alpha}}{\left(1-\cos p\pi\right)^{\alpha}} ,
\end{equation}
a bound that holds for \textit{every} $p$ and $\tau_r$, with no assumption on the size of $A$. Inserting \cref{integrand_bound} into $\mathcal{M}_{[0,\varepsilon]}$ and using the small-$p$ form \cref{C0_small_p} of the remaining trigonometric factors, we find
\begin{equation}\begin{split} \label{inner_region_estimate}
    \mathcal{M}_{[0,\varepsilon]} &\leq \frac{2\,\tau_r^{1-\alpha}}{1-\alpha}\int_0^{\varepsilon} dp\,
      \sin(p\pi x_0) \frac{\sin(p\pi)}{\left(1-\cos p\pi\right)^{1+\alpha}}
    \simeq \frac{2\,\tau_r^{1-\alpha}}{1-\alpha}\,\frac{2^{1+\alpha}x_0}{\pi^{2\alpha}}
      \int_0^{\varepsilon}\frac{dp}{p^{2\alpha}} = \\
    &= \frac{2^{2+\alpha}x_0}{\pi^{2\alpha}(1-\alpha)(1-2\alpha)}\;\varepsilon^{\,1-2\alpha}\,\tau_r^{1-\alpha}.
\end{split}\end{equation}
Since $\alpha<\frac{1}{2}$ the exponent $1-2\alpha$ is positive, and we therefore have
\begin{equation} \label{inner_region_vanishes}
    \frac{\mathcal{M}_{[0,\varepsilon]}}{\tau_r^{1-\alpha}} \sim \varepsilon^{\,1-2\alpha}  \sim  \tau_r^{\frac{2\alpha-1}{\beta}}
    \xrightarrow[\ \tau_r\shortrightarrow \infty\ ]{} 0 .
\end{equation}
The contribution of the interval $[0,\varepsilon]$ is thus a vanishing fraction of $\tau_r^{1-\alpha}$
and may be discarded. 

\textit{The interval $[\varepsilon,1]$.} Here requirement (ii) allows us to carry out the
approximation we were after. We begin with the inner integral $I_2$, substituting
$u\equiv \left(1-\cos p\pi\right)z$,
\begin{equation} \label{I2_substitution}
    I_2 = \int_0^{\tau_r}\frac{dz}{1+\left[\left(1-\cos p\pi\right)z\right]^\alpha}
    = \frac{1}{1-\cos p\pi}\int_0^{\left(1-\cos p\pi\right)\tau_r}\frac{du}{1+u^\alpha} .
\end{equation}
By requirement (ii) the integral over $u$ is dominated by the region $u\gg 1$, where
$1+u^\alpha\simeq u^\alpha$. Hence
\begin{equation}\begin{split} \label{I2_asymptotic}
    I_2 &\simeq \frac{1}{1-\cos p\pi}\int_0^{\left(1-\cos p\pi\right)\tau_r}\frac{du}{u^\alpha}
    = \frac{1}{1-\cos p\pi}\,\frac{\left[\left(1-\cos p\pi\right)\tau_r\right]^{1-\alpha}}{1-\alpha} =\\
    &=\frac{1}{1-\alpha}\,\frac{\tau_r^{1-\alpha}}{\left(1-\cos p\pi\right)^{\alpha}} .
\end{split}\end{equation}
The remaining $\tau_r$-dependent factor of $I_1$ simplifies as well: since $A\gg 1$ on this interval,
we have $\frac{1+A}{A}\simeq 1$. Putting the two together,
\begin{equation}\begin{split} \label{outer_region}
    \mathcal{M}_{[\varepsilon,1]} &= \int_{\varepsilon}^1 dp\, I_1\; I_2 \simeq \int_{\varepsilon}^1 dp\, \sin(p\pi x_0)\frac{\sin(p\pi)}{1-\cos(p\pi)}\cdot 1\cdot
      \frac{1}{1-\alpha}\frac{\tau_r^{1-\alpha}}{\left(1-\cos p\pi\right)^{\alpha}} = \\
    &=\frac{\tau_r^{1-\alpha}}{1-\alpha} \int_{\varepsilon}^1 dp\, \sin(p\pi x_0)
      \frac{\sin(p\pi)}{\left(1-\cos p\pi\right)^{1+\alpha}} .
\end{split}\end{equation}
The integral that remains resembles $C_0$ in \cref{C0_definition}, but for its lower
boundary. The missing sliver is small for exactly the reason given in requirement (i): by the estimate
already performed in\cref{inner_region_estimate},
\begin{equation} \label{C0_restored}
    \int_{\varepsilon}^1  dp\, \sin(p\pi x_0)\frac{\sin(p\pi)}{\left(1-\cos p\pi\right)^{1+\alpha}}
    = C_0\left[1+\mathcal{O}\!\left(\varepsilon^{\,1-2\alpha}\right)\right] .
\end{equation}
We may therefore restore the lower integration boundary to zero, and obtain
\begin{equation} \label{outer_region_final}
    \mathcal{M}_{[\varepsilon,1]} \simeq \frac{\tau_r^{1-\alpha}}{1-\alpha}
    \int_{0}^1 dp\, \sin(p\pi x_0)\frac{\sin(p\pi)}{\left(1-\cos p\pi\right)^{1+\alpha}}
    = \frac{C_0}{1-\alpha}\,\tau_r^{1-\alpha} .
\end{equation}

\noindent Combining \cref{inner_region_vanishes} with \cref{outer_region_final}, the first part of the split is negligible and the second gives the result,
\begin{equation} \label{MFPT_Large_Taur_PowerLaw}
    \text{MFPT}(\alpha<\tfrac{1}{2},\tau_r\gg 1, x_0) =\underbrace{\mathcal{M}_{[0,\varepsilon]}}_{\sim\,\tau_r^{\frac{1-2\alpha}{\beta}}\tau_r^{1-\alpha}}+\underbrace{\mathcal{M}_{[\varepsilon,1]}}_{\sim\,\tau_r^{1-\alpha}}
    \simeq \frac{C_0}{1-\alpha}\,\tau_r^{1-\alpha}.
\end{equation}
As promised, the cutoff $\varepsilon$ has dropped out of the final answer. It served only to separate the region where the approximation $A\gg 1$ fails from the region where it holds, and any $\varepsilon$ obeying requirements (i) and (ii) does the job equally well.

Collecting \cref{MFPT_Small_Taur_PowerLaw} and \cref{MFPT_Large_Taur_PowerLaw}, we summarize that the MFPT for $\tau_r\gg 1$ and $\tau_r\ll 1$ (at constant $\alpha<\frac{1}{2}$) is, as expressed in Eq.~\eqref{eq:MFPT_Asym},
\begin{equation} \label{MFPT_Large_Small_Taur_PowerLaw_Summary}
    \text{MFPT}(\alpha<\tfrac{1}{2},\tau_r, x_0) \simeq \left\{ \begin{array}{cc}
         \frac{1}{1-\alpha}\tau_r^{1-\alpha} \int_0^1 dp \sin(p\pi x_0)  \frac{\sin(p\pi)}{(1-\cos p\pi)^{1+\alpha}},&  \tau_r\gg 1, \\[5pt]
         \tau_r^{1-\alpha} \int_0^1 dp \sin(p\pi x_0)  \frac{\sin(p\pi)}{(1-\cos p\pi)^{1+\alpha}},& \tau_r\ll 1,
    \end{array}\right. =\left\{ \begin{array}{cc}
         \frac{C_0}{1-\alpha} \cdot \tau_r^{1-\alpha},&  \tau_r\gg 1, \\[5pt]
         C_0  \cdot \tau_r^{1-\alpha},& \tau_r\ll 1 .
    \end{array}\right.
\end{equation}
Both limits share the same power law $\tau_r^{1-\alpha}$ and differ only by the factor $\frac{1}{1-\alpha}$. The origin of this factor is transparent from the derivation: for $\tau_r\ll 1$ the inner integral $I_2$ is simply $\tau_r$, since its integrand stays close to unity over the whole interval, whereas for $\tau_r\gg 1$ the integrand decays as $z^{-\alpha}$ across most of the interval, which upon integration produces the extra $\frac{1}{1-\alpha}$ in \cref{I2_asymptotic}.

\begin{figure}[t]
\centering
\includegraphics[width=\textwidth]{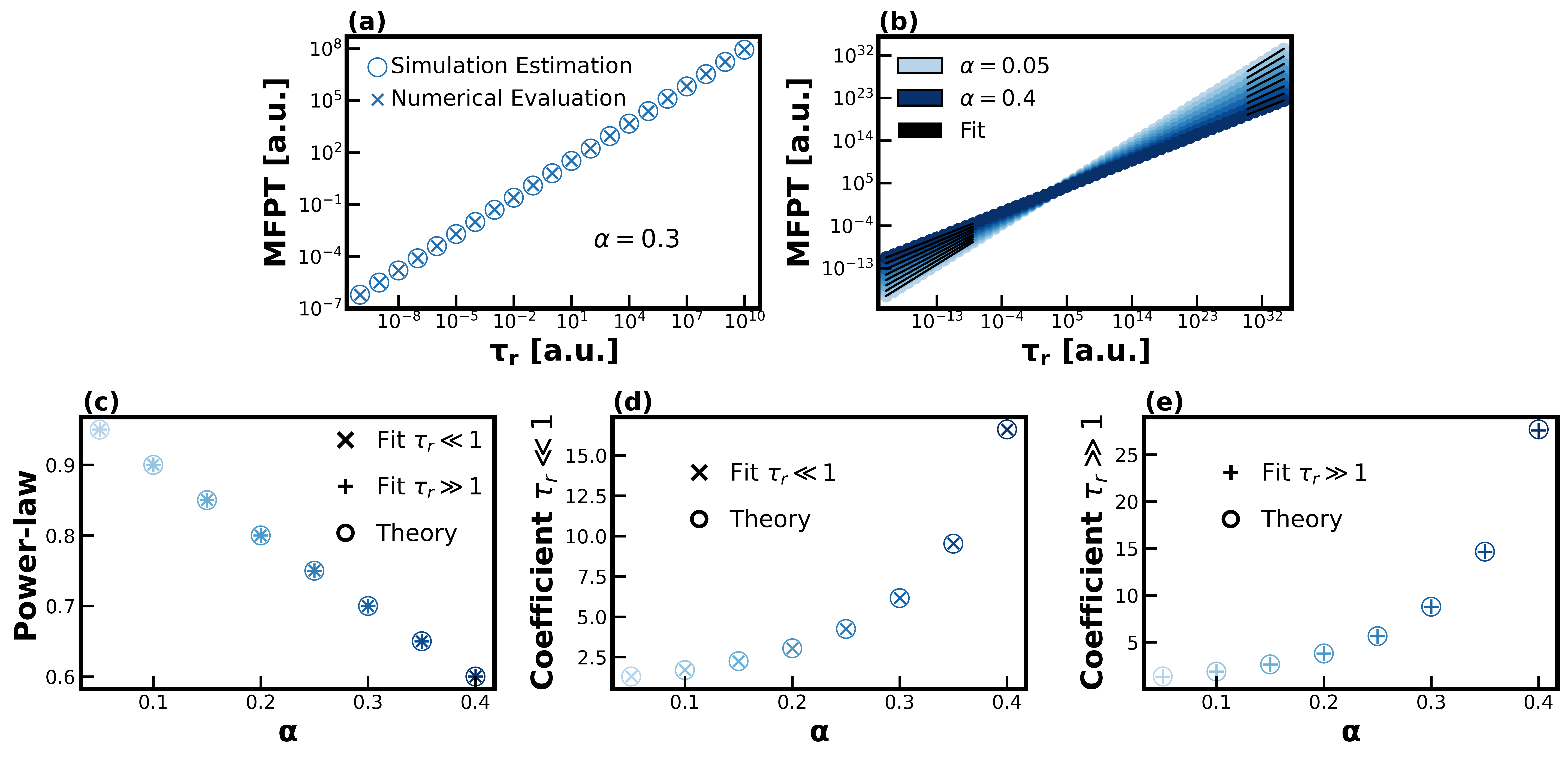}
  \caption{\textbf{Scaling of the MFPT with the Relaxation Time $\tau_r$.} \textbf{a.} MFPT as a function of the relaxation time $\tau_r$ for $\alpha=0.3$, simulation estimation and numerical evaluation are shown in circles and x-markers respectively. \textbf{b.} Numerical evaluations of the MFPT in \cref{eq:final_MFPT} as a function of $\tau_r$ for multiple values of $\alpha$, from $\alpha=0.05$ (lightest blue) to $\alpha=0.4$ (darkest blue). Fits for both $\tau_r\ll 1$ and $\tau_r\gg 1$ are in solid lines. \textbf{c.} Power-law as a function of $\alpha$. The power-law is from MFPT vs. $\tau_r$, for $\tau_r\gg 1$ (x markers) and $\tau_r\ll 1$ (+ markers). The theoretical power-law for both ranges is $1-\alpha$ and is denoted by empty circles. \textbf{d.} Coefficient for $\tau_r\ll 1$ as a function of $\alpha$. The coefficients from the power-law of MFPT vs. $\tau_r$, for $\tau_r\ll 1$ regime. Coefficient from the fit, denoted by x markers, and theory in empty circles. \textbf{e.} Coefficient for $\tau_r\gg 1$ as a function of $\alpha$. The coefficients from the power-law of MFPT vs. $\tau_r$, for $\tau_r\gg 1$ regime. Coefficient from the fit denoted by + markers and theory by empty circles.   }  \label{Supporting_MFPT}
\end{figure}

\cref{Supporting_MFPT} shows a comparison and fit analysis of the behavior of the MFPT as a function of the relaxation time $\tau_r$. In \cref{Supporting_MFPT}(a), we show the MFPT as a function of $\tau_r$, and compare the simulation estimation (empty circle) with the numerical evaluation (x marker) for $\alpha = 0.3$. The two appear to agree over several orders of magnitude in $\tau_r$. In \cref{Supporting_MFPT}(b), we plot the numerical estimation of the MFPT as a function of $\tau_r$, for multiple values of $\alpha$, from $\alpha = 0.05$ (lightest blue) to $\alpha = 0.4$ (darkest blue). The fit for $\tau_r\ll 1$ and $\tau_r\gg 1$ are indicated by black lines. We show the result of those fits, both in terms of the slope, the power-law, and in terms of the coefficient in \cref{Supporting_MFPT}(c)-(e). The power-law for both regimes, $\tau_r\ll 1$ and $\tau_r\gg 1$, are identical and equal $1-\alpha$, \cref{MFPT_Large_Small_Taur_PowerLaw_Summary}. Indeed, as shown in \cref{Supporting_MFPT}(c), the power-law as a function of $\alpha$, we see an accurate match between the theoretical $1-\alpha$ (empty circles) and the fit results (x markers for $\tau_r\ll 1$ and + markers for $\tau_r\gg 1$). Lastly, we check whether the coefficient of the power-law of the fit matches the theoretical value. The coefficient of $\tau_r\ll 1$ and $\tau_r\gg 1$ are almost identical, differing by a factor of $\frac{1}{1-\alpha}$; for $\tau_r\ll 1$, the theoretical coefficient is the integral $C_0\equiv \int_0^1 dp \sin(p\pi x_0)  \frac{\sin(p\pi)}{(1-\cos p\pi)^{1+\alpha}}$, and for $\tau_r\gg 1$ the theoretical coefficient is $\frac{C_0}{1-\alpha}$, \cref{MFPT_Large_Small_Taur_PowerLaw_Summary}. In \cref{Supporting_MFPT}(d)-(e) we compare the theoretical and fitted coefficients for $\tau_r\ll 1$ and $\tau_r\gg 1$ respectively. For both plots, the theoretical value is denoted by an empty circle, whereas the fitted result for $\tau_r\ll 1$ and $\tau_r\gg 1$ is shown in x and + markers, respectively. In both cases we see a clear match between the fitted and theoretical values, across all values of $\alpha$.

\bibliographystyle{unsrt}
\bibliography{Ref}

\end{document}